\documentclass[journal,onecolumn,draftclsnofoot,english,12pt]{IEEEtran}
\usepackage[T1]{fontenc}
\usepackage{float}
\usepackage{amsmath}
\usepackage{dsfont}
\usepackage{amsthm}
\usepackage{amssymb}
\usepackage{stackrel}
\usepackage{graphicx}
\usepackage{setspace}
\usepackage{url}
\usepackage{color}

\usepackage{caption}
\usepackage{subfigure}
\usepackage{algorithm}
\usepackage{algorithmic}

\makeatletter

\floatstyle{ruled}
\newfloat{algorithm}{tbp}{loa}
\providecommand{\algorithmname}{Algorithm}
\floatname{algorithm}{\protect\algorithmname}

\IEEEoverridecommandlockouts
\usepackage{ifpdf}
\usepackage{cite}
\ifCLASSOPTIONcompsoc
\usepackage[caption=false,font=normalsize,labelfont=sf,textfont=sf]{subfig}
\else
\usepackage[caption=false,font=footnotesize]{subfig}
\fi
\usepackage{amsmath}
\usepackage{mathtools}
\usepackage{booktabs} 
\usepackage{multirow}
\usepackage{setspace}
\usepackage{bm}
\usepackage{mathrsfs}

\usepackage{bbm}

\@ifundefined{showcaptionsetup}{}{%
 \PassOptionsToPackage{caption=false}{subfig}}
\usepackage{subfig}
\makeatother

\usepackage{color}

\allowdisplaybreaks[3]

\usepackage{caption}
\usepackage{subfigure}
\usepackage{url}

\usepackage[dvipsnames]{xcolor}
\usepackage{soul}  
\usepackage{soul}
\soulregister\cite7 
\soulregister\citep7 
\soulregister\citet7
\soulregister\ref7
\soulregister\pageref7
\usepackage{colortbl}
\definecolor{mygray}{gray}{.9}
\definecolor{mygray1}{gray}{1.}
\definecolor{mypink}{rgb}{.99,.91,.95}
\definecolor{mycyan}{cmyk}{.3,0,0,0}
\usepackage{tabu} 
\usepackage{array}  
\usepackage{adjustbox}
\usepackage{makecell}

\usepackage{ulem} 
\usepackage{babel}

\begin{document}

\captionsetup[figure]{font={small}, name={Fig.}, labelsep=period}
\newcommand{\tabincell}[2]{\begin{tabular}{@{}#1@{}}#2\end{tabular}}
\title{
Digital Twin-Aided Learning for Managing Reconfigurable Intelligent Surface-Assisted, Uplink, User-Centric Cell-Free Systems

}

\author{Yingping Cui, Tiejun Lv,~\IEEEmembership{Senior Member,~IEEE}, \\
Wei Ni,~\IEEEmembership{Senior Member,~IEEE}, and Abbas Jamalipour,~\IEEEmembership{Fellow,~IEEE} 

\thanks{Y. Cui and T. Lv  are with the School of Information and Communication Engineering, Beijing University of Posts and Telecommunications (BUPT), Beijing 100876, China (e-mail: \{cuiyingping, lvtiejun,\}@bupt.edu.cn).}
\thanks{W. Ni is with the Commonwealth Scientific and Industrial Research Organisation
(CSIRO), Sydney 2122, Australia (e-mail: Wei.Ni@data61.csiro.au).}
\thanks{A. Jamalipour is with the University of Sydney, NSW, Australia (e-mail: a.jamalipour@ieee.org).}
}

\maketitle

\begin{abstract}
This paper puts forth a new, reconfigurable intelligent surface (RIS)-assisted, uplink, user-centric cell-free (UCCF) system managed with the assistance of a digital twin (DT).  Specifically, we propose a novel learning framework that maximizes the sum-rate by jointly optimizing the access point and user association (AUA), power control, and RIS beamforming.
This problem is challenging and has never been addressed due to its prohibitively large and complex solution space.
Our framework decouples the AUA from the power control and RIS beamforming (PCRB) based on the different natures of their variables, hence reducing the solution space.
A new position-adaptive binary particle swarm optimization (PABPSO) method is designed for the AUA. Two twin-delayed deep deterministic policy gradient (TD3) models with new and refined state pre-processing layers are developed for the PCRB.
Another important aspect is that a DT is leveraged to train the learning framework with its replay of channel estimates stored. The AUA, power control, and RIS beamforming are only tested in the physical environment at the end of selected epochs.
Simulations show that using RISs contributes to considerable increases in the sum-rate of UCCF systems, and the DT dramatically reduces overhead with marginal performance loss. 
The proposed framework is superior to its alternatives in terms of sum-rate and convergence stability.

\end{abstract}


\IEEEpeerreviewmaketitle

\section{Introduction}
Cell-free systems provide a promising network architecture candidate for next-generation wireless networks~\cite{8097026}, due to its gain over existing cellular systems in diversity~\cite{7227028} and subsequently, throughput~\cite{7827017} or spectral efficiency (SE)~\cite{8886730}. Compared to the conventional cell-free systems where all access points (APs) serve every user equipment (UE) within their coverage, a more practical architecture is a user-centric cell-free (UCCF) system with each UE served only by its associated APs (as opposed to all APs).
Provided effective AP-UE association (AUA), a UCCF system would reduce the hardware and software complexity of the APs and the backhaul overhead of the APs~\cite{8000355,8901451}.

Reconfigurable intelligent surface (RIS) is an increasingly popular technique considered for the cost-effective deployment of next-generation wireless networks~\cite{8910627}.
Comprising a large number of passive reflective elements, an RIS can have the reflection coefficients of its elements individually adjusted to ameliorate wireless propagation environments and improve system capacity~\cite{9459505}, achievable rate~\cite{9115725}, or energy efficiency (EE)~\cite{9363171}. The consideration of an RIS is imperative for a future-proof design of wireless communication systems.

A new RIS-assisted UCCF system is promising, joining the benefits of both UCCF and RIS for future wireless networks.
The optimization of the system necessitates a holistic joint design of AUA, power control, and RIS passive beamforming, and is nontrivial due to a prohibitively large and complex solution space.
To the best of our knowledge, the RIS-assisted UCCF system has never been studied in the literature.
Conventional approaches relying generally on prior information lack flexibility~\cite{8790780} and can hardly ensure effectiveness and reliability. 
Learning-based methods can make decisions adaptively by fitting learning policies between the input and output to a nonlinear model~\cite{8714026}, offering the potential to effectively optimize the AUA, power control, and RIS beamforming of the RIS-assisted UCCF system.
 
{\color{black}Digital twin (DT) provides a new paradigm for effective management of complex systems, e.g., the next-generation wireless networks, by creating a real-time digital simulation model of the systems~\cite{8822494}.
It can monitor the operating state of the systems in real-time, record and replay the past experience of the systems, and predict future states or operations based the experience replay~\cite{9447819,8972134}.
A DT-based virtual environment can be particularly useful when direct interactions with the actual physical environment are practically unaffordable, e.g., for real-time resource allocation of RIS-assisted UCCF systems.
}

This paper presents a new learning-based framework that maximizes the sum-rate of the new RIS-assisted, uplink, UCCF system with the assistance of a DT. The AUA, power control, and RIS beamforming are jointly optimized by training the learning model with the DT that replays the channel estimates stored to emulate the rewards to drive the learning.

The key contributions of the paper are summarized as follows.
\begin{enumerate}
\item 
A new problem is formulated to maximize the sum-rate of the new RIS-assisted, uplink, UCCF system with only the imperfect channel state information (CSI) of the end-to-end channels between the UEs and APs (including those via the RISs).

\item
A new learning framework is proposed to solve the new problem by decoupling the AUA, power control, and RIS beamforming based on the different natures of their variables, while capturing their interplay through the design of interactive rewards. This considerably reduces the action space of the learning framework.

\item A new position-adaptive binary particle swarm optimization (PABPSO) method is designed for the AUA to eliminate infeasible particles that would be otherwise generated by the conventional BPSO.
Two twin delayed deep deterministic policy gradient (TD3) models are paralleled for the PCRB (i.e., power control and RIS beamforming) with refined state pre-processing, thereby improving training convergence and reducing complexity.

\item 
A DT is adopted to train the proposed deep reinforcement learning (DRL) model based its stored channel estimates and emulated rewards for learning in the virtual environment. Interactions with the physical environment may only take place once at the end of selected epochs to refine the DT-based training process. 
\end{enumerate}

Extensive simulations show that the proposed framework is superior to its possible alternatives in terms of achievable rate and robustness.
The use of the DT can dramatically cut off the overhead arising from actual transmissions in the physical environment and substantially reduce the overhead and delay. 
The significant gain of the RIS is also demonstrated. 

The rest of this paper is organized as follows. In Section \ref{section:relatedwork}, the related works are discussed. In Section \ref{section:systemmodel}, the RIS-assisted, uplink, UCCF system model is presented, and the problem is formulated. In Section \ref{section:problem_transformation}, we transform the problem into two subtasks. In Section~\ref{section:proposed_framework}, a new learning-based semi-parallel framework is proposed. Simulation results are presented in Section \ref{section:sim}, followed by conclusions in Section~\ref{section:con}. 

\begin{table*}
\begin{center}
\caption{Notation and Description.}
\begin{tabular}{l|l}

\Xhline{1pt}
\textbf{Notation}  & \textbf{Description} \\
\hline
$K$, $M$, $B$, $N$ & The numbers of UEs, APs, RISs, and elements of each RIS \\
$\mathcal M(k)$ & The set of APs collaboratively serving UE $k$ \\
$h_{mk}$ & The direct channel from AP $m$ to UE $k$\\
$\mathbf G_m$ & The channels between AP $m$ and all RISs\\
$\mathbf v_{k}$ & The channels between UEs $k$ and all RISs\\
$\theta_{bn}$, $\phi_{bn}$ & The reflection coefficient and phase shift of the $n$-th element of RIS $b$\\
$h_{mbnk}$ & The\,reflected\,channel\,from\,UE\,$k$\,to\,AP\,$m$\,via\,the\,$n$-th\,element\,of\,RIS\,$b$\\
$f_{mk}$ & The equivalent channel between UE $k$ and AP $m$\\
$\mathcal F_c$ & The feasible set of the reflection coefficients\\
{\small{$\bar h_{mk}$, $\tilde h_{mk}$}} & The LoS and non-LoS components of $h_{mk}$\\
{\small{$\bar h_{mbnk}$, $\tilde h_{mbnk}$}} & The LoS and non-LoS components of $h_{mbnk}$\\
{\small{$\hat h_{mk}$, $\hat h_{mbnk}$, $\hat f_{mk}$}} & The channel estimates of $h_{mk}$, $h_{mbnk}$ and $f_{mk}$, respectively\\
$p_k$ & The transmit power of UE $k$\\
$p_{k,\mathrm{max}}$ & The maximum transmit power of UE $k$\\
$\sigma_{m}^2$ & The variance of the noise at AP $m$\\
$\alpha_k$ & The SINR of UE $k$\\
$\mathcal R_k$ & The achievable data rate of UE $k$\\
$\mathcal R_{k,\mathrm{min}}$ & The minimum data rate of UE $k$\\
$\boldsymbol{\Lambda}$ & The AUA matrix\\
$Z$ & The number of epochs\\
$Y$ & The number of episodes of the PCRB per epoch\\
$T$ & The total step number of the PCRB module\\
$L$ & The number of iterations of the AUA per epoch\\
$I$ & The number of particles\\
$\mathbf x_i$, $\mathbf v_i$ & The position and velocity of particle $i$\\
$f(\mathbf x_{i})$ & The fitness value of $\mathbf x_{i}$\\
$\mathbf{\dot p}_i$ & The local best solution of particle $i$\\
$\mathbf{\dot g}$ & The global best solution of all particles\\
$\mathbf{p}^{\text{AUA}}$ & The power control vector input to the AUA module\\
$\boldsymbol{\phi}^{\text{AUA}}$ & The\,matrix\,collecting\,the\,phase\,shifts\,of\,all\,RISs\,input\,to\,the\,AUA\,module\\
$\boldsymbol a_t^{\mathrm p}$ & The actions output by the PC agent\\
$\boldsymbol{a}_t^{\mathrm R}$ & The actions output by the RB agent\\
$r_t$ & The immediate reward of the PCRB module at step $t$\\
$\boldsymbol s_t^{\text{total}}$ & The environment state at step $t$\\
$\boldsymbol s_t^{\mathrm {o}}$ & The infrequently changing part of environment state at step $t$\\
$\boldsymbol s_t^{\mathrm {d}}$ & The frequently changing part of environment state at step $t$\\
$\boldsymbol s_t$ & The state observed by the PC and RB agents at step $t$\\
$\zeta_{i}^{\mathrm {p}}$ & The parameters of the $i$-th evaluated-critic-network for the PC agent\\
$\zeta_{i}^{\mathrm {p'}}$ & The parameters of the $i$-th target-critic-network for the PC agent\\
$\mu^{\mathrm R}$ & The parameters of the evaluated-actor-network for the RB agent\\
$\mu^{\mathrm R'}$ & The parameters of the target-actor-network for the RB agent\\
\Xhline{1pt}
\end{tabular}
\end{center}
\end{table*}

\emph{Notations}: $x$ stands for a scalar, $\mathbf x$ represents a vector, and $\mathbf X$ denotes a matrix; $x_{ij}$ represents the $(i,j)$-th entry of a matrix $\mathbf X$;
$\text{diag}(\cdot)$ denotes the diagonalization;
$(\cdot)^{\ast}$, $(\cdot)^{\mathsf{T}}$, and $(\cdot)^{\mathsf{H}}$ represent the conjugate, the transpose, and the conjugate transpose;   $\jmath=\sqrt{-1}$, denotes the imaginary unit; $\mathbb{P} [\cdot]$ and $\mathbb{E} [\cdot]$ denote probability and expectation, respectively. $\mathbf Y=\text{Reshape}(\mathbf X,m,n)$ denotes the matrix transformation that returns an $m\times n$ matrix $\mathbf Y$, where the elements of $\mathbf Y$ are obtained from $\mathbf X$ by column. 
Notations used are collected in Table {\uppercase\expandafter{\romannumeral1}}.

\section{Related Work}\label{section:relatedwork}

\subsection{Digital Twins for Wireless Network}
DTs have been increasingly considered in wireless systems to leverage its function of visualization and replay.
Luan \emph{et al.} \cite{luan2021paradigm} described a basic DT-assisted communication model, where a DT served as a simulation platform for testing wireless system designs.
Sun \emph{et al.} \cite{9351542} designed a DT of aerial-assisted internet of vehicles to record vehicle preference information and assist in resource management based on the alternating direction method of multipliers (ADMM).
Sheen \emph{et al.} \cite{sheen2020digital} presented a DT framework for RIS-assisted wireless networks to learn the mapping function between RIS beamforming and the receiver locations and maximize the downlink sum-rate. The DT recorded the labeled data for training the mapping function. Deng \emph{et al.} \cite{9420037} combined expert knowledge, reinforcement learning (RL) and DT, where the DT acted as a virtual environment to interact with RL and select actions between the output of the RL and expert knowledge. 
Liu \emph{et al.} \cite{9447819} proposed a DT-assisted task offloading scheme for mobile-edge computing, where users selected their collaborative servers with the help of a DT storing global system data (e.g., CSI) and used DRL to optimize their task offloading strategy. 
Lv \emph{et al.} \cite{lv2021beyond} deployed a deep learning (DL) algorithm with the support of DT to estimate the channels of an unmanned aerial vehicle, and investigated the performance differences between the DL-based method and other channel estimation methods. 
However, these designs cannot solve the problem studied in this paper, because they cannot address the AUA that is closely coupled with power control and RIS beamforming in the problem. 

\subsection{Resource Management in UCCF Systems}
Power control and AUA have been studied in UCCF systems with no RIS.
Buzzi \emph{et al.}~\cite{8901451} proposed two power allocation strategies to maximize the sum-rate and the minimum rate by using alternating optimization and sequential convex programming. The results showed that a UCCF system generally outperforms its cell-free counterpart, especially in the uplink. Alonzo \emph{et al.} \cite{8676377} proposed a low-complexity power allocation strategy to maximize the global EE of UCCF systems.
No AUA was considered in \cite{8901451} and \cite{8676377}.
Liu \emph{et al.} \cite{7676391} jointly optimized AUA, power control, and user-side interference cancellation to maximize the spectral- and energy-efficiency of a downlink UCCF system. A generalized weighted minimum mean square error algorithm was used to optimize the power control. An exhaustive search (ES) was used to optimize the AUA.
D'Andrea \emph{et al.} \cite{9443536} designed a position-based virtual clustering approach for AUA to maximize the sum-rate. 
Ngo \emph{et al.} \cite{8097026} utilized sequential convex approximation to solve downlink power control, and designed two low-complexity AUA algorithms based on the received power and large-scale fading information.
These AUA techniques are not applicable in the presence of~RISs.

While the gain of RISs has been demonstrated in the capacity and EE of cell-free systems~\cite{9363171,9459505,9448858}, no RIS has been incorporated in UCCF systems; in other words, no AUA has been considered in those studies.
Bashar \emph{et al.} \cite{9322151} designed power control and RIS beamforming to maximize the minimum rate in the downlink of an RIS-assisted cell-free system utilizing geometric programming and semidefinite programming.
To maximize a weighted sum-rate of an RIS-assisted, downlink, cell-free system, 
Zhang \emph{et al.} \cite{9459505} proposed a centralized beamforming scheme to jointly optimize the beamforming at the APs and RISs, and showed that the use of an RIS contributes to the capacity of cell-free systems. 
Huang \emph{et al.} \cite{9298843} designed a decentralized AP and RIS beamforming scheme based on the ADMM, where each AP decides its beamforming locally. 
The design relieved the computational burden on the edge server, compared to its centralized counterpart developed in \cite{9459505}. 
Le \emph{et al.} \cite{9363171} developed a low-complexity alternating descent-based iterative algorithm to maximize the EE of an RIS-assisted, downlink, cell-free system iteratively, and obtained locally optimal analog beamforming at the APs and RISs.
Zhang \emph{et al.} \cite{9352948} designed a hybrid downlink beamforming scheme consisting of digital beamforming at the BSs and analog beamforming at the RIS to maximize EE iteratively. 
Unlike the above studies of downlink systems, Liu \emph{et al.}~\cite{9448858} considered an uplink RIS-assisted cell-free system and employed Lagrangian transform and fractional programming to maximize the worst-case~EE.

DRL has been applied to cell-free or UCCF systems with no RIS~\cite{9483914,2020distributedDDPG,9174775} or RIS-assisted conventional cellular systems~\cite{8968350,9110869,9473585}.
Zhao \emph{et al.}~\cite{9483914} used two DRL models, i.e., deep Q-networks (DQN) and deep deterministic policy gradient (DDPG), to learn the power control that maximizes the SE of a cell-free system.
Fredj \emph{et al.} \cite{2020distributedDDPG} developed a distributed DDPG algorithm to design the beamforming at the APs in a cell-free system. 
Al-Eryani \emph{et al.} \cite{9174775} jointly optimized AUA and beamforming at the APs to maximize the sum-rate of a downlink UCCF system. A hybrid DRL framework was designed, where a DQN learned the AUA and DDPG produced the beamforming. 
Feng \emph{et al.} \cite{8968350} investigated the RIS beamforming for an RIS-assisted, downlink, cellular system, and applied DRL to tackle the non-convex unit modulus constraints of the RIS phase shifts.
Huang \emph{et al.} \cite{9110869} proposed a joint design of active and passive beamforming in an RIS-assisted cellular system using DRL.
Only a single RIS was considered in \cite{8968350} and \cite{9110869}.
Kim \emph{et al.} \cite{9473585} considered a multi-cell system with the assistance of multiple RISs, and proposed a joint control scheme based on a multi-agent DQN to design the discretized transmit power control, RIS beamforming, and BS combining.
All these algorithms require direct interactions with the physical environments to test the learned results. None has considered the potential use of DTs. 

\section{System Model}\label{section:systemmodel}
In this section, we first introduce the considered system architecture, followed by the wireless channel model. Finally, we present the achievable sum-rate.

\subsection{System Architecture}

{\color{black}We consider a new DT-empowered, uplink UCCF system,}
where $K$ number of single-antenna UEs are served cooperatively by $M$ number of single-antenna APs ($K \leq M$) with the assistance of $B$ number of RISs. The UEs and APs are randomly uniformly distributed within the considered area. Each RIS is an $N$-element uniform rectangular planar array. 
The APs, UEs, and RISs are collected by $\mathcal{M}=\{1,\cdots,M\}$, $\mathcal{K}=\{1, \cdots,K\}$, and $\mathcal{B}=\{1, \cdots,B\}$, respectively.

Consider a time-division duplexing (TDD) protocol as done in \cite{8901451,8676377}, where a coherence time is divided between downlink and uplink. In the downlink, the APs precode and transmit data based on the CSI estimated in the preceding uplink under the assumption of channel reciprocity. In the uplink, the UEs transmit signals to the APs following a user-centric approach.  
{\color{black}
For illustration convenience, we consider a single channel in which each UE can be served by multiple APs while an AP can only serve one UE at a time, as illustrated in Fig.~\ref{fig:J_global}. The set of APs collaboratively serving UE $k$ is denoted by $\mathcal{M}(k)$.}
Each AP detects signals from its associated UE.
All APs are connected via fronthaul to an edge server, at which the detected signals are combined and decoded to recover the UEs' data.
Toward the end of the uplink stage, the UEs send pilot signals, based on which the APs estimate the CSI for the next downlink transmissions.

A DT is employed to serve as an intermediate layer (and an agile and manageable interface) between the hardware and software, 
e.g., the APs and server, and the proposed control algorithm solving the AUA, power control, and RIS beamforming. 
On the one hand, the DT can instruct the hardware to implement the control decisions made by the algorithm into the physical environment, and test the response of the environment to the decisions, e.g., the achievable data rates of the UEs~\cite{10.1117/12.2618612}. 
On the other hand, the DT can record and replay the parameters measured in the physical environment, e.g., channel estimates~\cite{lv2021beyond}. It can emulate the possible environmental response to assess the control decisions virtually when the actual responses are too expensive to acquire (and would incur prohibitively frequent interactions between the environment and the control algorithm and unacceptable delays). In this paper, we propose a new learning algorithm to optimize AUA, power control, and RIS beamforming in the considered RIS-assisted, UCCF system, as will be discussed later in Section~\ref{section:proposed_framework}. 
The use of a DT significantly reduces direct interactions between the algorithm and the physical environment, hence facilitating the implementation of the algorithm.

\subsection{Channel Model}
Let $h_{mk} \in \mathbb{C}^{1 \times 1}$ denote the direct channel from AP $m$ to UE $k$, $\mathbf{g}_{mb} \in \mathbb{C}^{N \times 1}$ denote the channel from RIS $b$ to AP $m$, and $\mathbf{v}_{bk} \in \mathbb{C}^{N \times 1}$ denote the channel from UE $k$ to RIS $b$. The RISs have continuous phase shifts \cite{9115725}.
${\theta}_{bn} \in \mathcal F_{\mathrm {c}}$ is the reflection coefficient of the $n$-th reflection element of RIS $b$, where
$\mathcal F_c=\{\theta_{bn} = e^{\jmath \phi_{bn}} \vert \phi_{bn} \in [0, 2\pi), \forall b,n\}$ is the set of feasible reflection coefficients. 
Let $\boldsymbol{\theta}_b=[{\theta}_{b1},\cdots,{\theta}_{bN}]^{\mathrm{T}} \in \mathbb{C}^{N \times 1}$ with ${\theta}_{bn}=e^{\jmath\phi_{bn}}$.
$\boldsymbol{\theta}=[\boldsymbol{\theta}_1^{\mathrm{T}},\cdots,\boldsymbol{\theta}_B^{\mathrm{T}}]^{\mathrm{T}}\in \mathbb{C}^{BN \times 1}$ collects the reflection coefficients of all RISs. 
The equivalent channel between UE $k$ and AP $m$ is given by
\begin{align}
f_{mk}&=h_{mk}+\sum_{b=1}^B \mathbf{g}_{mb}^{\mathrm {H}} \boldsymbol {\Theta}_b \mathbf{v}_{bk}=h_{mk}+ \boldsymbol {\theta}^{\mathrm {H}} (\mathbf G_{m}^{\mathrm {H}} \mathbf {v}_k) \nonumber \\ 
&=h_{mk}+ \sum_{b=1}^B\sum_{n=1}^{N}g_{mbn}^{*} v_{bnk} e^{\jmath\phi_{bn}}=h_{mk}+ \sum_{b=1}^B\sum_{n=1}^{N}h_{mbnk} e^{\jmath\phi_{bn}},
\label{eq:fmk}
\end{align}
where $\boldsymbol{\Theta}_b=\text{diag}(\boldsymbol{\theta}_b) \in \mathbb{C}^{N \times N}$ is the diagonal matrix of the reflection coefficients of RIS $b$; $\mathbf{G}_m=\text{diag}(\mathbf{g}_{m1},  \cdots, \mathbf{g}_{mB})\in \mathbb{C}^{BN \times BN}$ is the diagonal matrix of the channels between all RISs and AP $m$;
$\mathbf{v}_k=[\mathbf{v}_{1k}^{\mathrm T}, \cdots , \mathbf{v}_{Bk}^{\mathrm T}]^{\mathrm{T}} \in \mathbb{C}^{BN \times 1}$ collects the channels between UE $k$ and all RISs;
and $h_{mbnk}=g_{mbn}^{*} v_{bnk}$ is the RIS-reflected channel from UE $k$ to AP $m$ via the $n$-th element of RIS $b$. Here, $g_{mbn}$ is the channel from the $n$-th element of RIS $b$ to AP $m$, and $v_{bnk}$ is the channel from UE $k$ to the $n$-th element of RIS $b$.

According to (\ref{eq:fmk}),
the estimate of $f_{mk},\,\forall m,k$ can be given by
\begin{align}
    \hat f_{mk}=\hat h_{mk}+ \sum_{b=1}^B\sum_{n=1}^{N}\hat h_{mbnk} e^{\jmath\phi_{bn}},
\label{eq:hat_fmk}
\end{align}
where $\hat h_{mk}$ and $\hat h_{mbnk}$ are the estimates of $h_{mk}$ and $h_{mbnk}$, respectively. 

In this paper, only the end-to-end channels between any UE $k$ and AP $m$, including the direct channel, i.e., $h_{mk}$, and the RIS-cascaded channel, i.e., $h_{mbnk}$, are needed to be estimated.
No individual channel to or from an RIS, i.e., $g_{mbn}$ and $v_{bnk}$, is needed.
Estimating the individual channels to or from an RIS is generally challenging~\cite{9732214}. The use of only the end-to-end channels can substantially facilitate practical implementations.

The channels are estimated {\color{black}with the aid of the pairwise uplink orthogonal unit pilot sequences from the $K$ UEs.} An MMSE estimator can be applied to estimate
$\hat h_{mk}$ and $\hat h_{mbnk}$~\cite{8645336}:
\begin{equation}
\hat h_{mk}=\bar h_{mk}^{\mathrm{E}}+\frac{\sqrt {p_k^\text{u}}\beta_{mk}}{p_k^\text{u}\beta_{mk}+\sigma_{m}^2}(\sqrt{p_k^\text{u}\beta_{mk}} \tilde h_{mk}+\check u_{mk}^{p});
\label{eq:hathmk}
\end{equation}
\begin{equation}
\hat h_{mbnk}=\bar h_{mbnk}^{\mathrm{E}}+\frac{\sqrt{p_k^\text{u}}\beta_{mbk}} {p_k^\text{u}\beta_{mbk}+\sigma_{m}^2}(\sqrt{p_k^\text{u}\beta_{mbk}}\tilde h_{mbnk}+\check u_{mbnk}^{\mathrm p}),
\label{eq:hathmbnk}
\end{equation}
where 
$p_k^\text{u}$ denotes the power of the uplink pilot sent by UE $k$;
$\check{u}_{mk}^{\mathrm p} \sim\mathcal{CN}(0,\sigma_{m}^2)$ and $\check{u}_{mbnk}^{\mathrm p} \sim\mathcal{CN}(0,\sigma_{m}^2)$
are the additive white Gaussian noises (AWGNs) at AP $m$;
$\beta_{mk}$ and $\beta_{mbk}$ are two parameters that depend on the distance-dependent large-scale path loss and Rician fading factors;
and 
$\bar h_{mk}^{\mathrm{E}}$ and $\bar h_{mbnk}^{\mathrm{E}}$ are the equivalent LoS components of $h_{mk}$ and $h_{mbnk}$, respectively.
According to~\cite{dai2022two}, $\beta_{mk}$, $\beta_{mbk}$, $\bar h_{mk}^{\mathrm{E}}$, and $\bar h_{mbnk}^{\mathrm{E}}$ could be invariant for a long period and can be treated as deterministic.

The DT records the historical data of the channel estimates between the UEs at different locations and APs. The records of $\hat h_{mk}$ and $\hat h_{mbnk}$, $\forall m, b, n, k$ can emulate the physical environment for training DRL models, thereby substantially cutting off the overhead.
Once in a while, the DRL models are deployed or tested in the physical environment. The deviation of the models trained against the DT in the physical environment can contribute to the subsequent training of the models in the DT. 
When testing the models in the physical environment, 
the accuracy of the estimated channels recorded in the DT can be improved by incorporating the new physical measurements. 

\subsection{Uplink Data Transmission}
Let $s_k$ denote the transmit symbol of UE $k$, and $p_k$ denote the transmit power. The received signal at AP $m$ is given by
\begin{align}
y_{m}=\sum\limits_{\begin{smallmatrix}
 {k}=1
\end{smallmatrix}}^{K}{\sqrt{p_k}f_{mk}s_k}
+u_{m},
\end{align}
where $u_m \sim \mathcal{CN}(0,\sigma_{m}^2)$ is the AWGN at AP $m$.
Each AP only decodes its served UE and treats the other UEs' signals as interference in the UCCF system.
The low-complexity matched filtering is applied, so that each AP can detect the signals only using its local estimated CSI \cite{7827017}.
Suppose that AP $m$ serves UE $k$, i.e., by multiplying the received signal $y_m$ with the conjugate of its estimated CSI denoted by $\hat f_{mk}^*$, and
$\hat f_{mk}^*=\hat h_{mk}^*+\sum_{b=1}^B \sum_{n=1}^N \hat h_{mbnk}^*e^{-\jmath\phi_{bn}}$. 
The result ${\check y_{mk}} = \hat f_{mk}^ * {y_{{m}}}$ is sent to the server, where the signals of UE $k$ detected by the set of APs, $\mathcal{M}(k)$, are combined:
\begin{align}
\hat{x}_k =&\sum_{m\in \mathcal M(k)} \check{y}_{mk}=\sum_{m=1}^M \lambda_{km} \check{y}_{mk} \nonumber \\
=&\sqrt{p_k}\sum_{m=1}^M \lambda_{km}\hat {f}_{mk}^* f_{mk}s_k +\sum_{m=1}^M \lambda_{km}\hat {f}_{mk}^*u_m
+\sum_{m=1}^M \lambda_{km}\sum\limits_{{k}^{\prime}=1,{k}^{\prime}\ne k}^{K}\sqrt{p_{{k}^{\prime}}}\hat {f}_{mk}^*{f_{m{k}^{\prime}}s_{k'}}.
\end{align}
Here, $\lambda_{km} \in \{0,1\}, \, \forall k , m $. $\lambda_{km} = 1$, if AP $m$ is associated with UE $k$; or $\lambda_{km}=0$, otherwise.

As a result, the SINR of UE $k$ is
${{\alpha}_{k}}=\frac{{{p}_{k}}{{\left| \sum\limits_{m=1}^{M}{{{\lambda}_{km}} \hat{f}_{mk}^{*}{{f}_{mk}}}\right|}^{2}}}{\sum\limits_{m=1}^{M}{{{\sigma_{m}^{2}\left| {{\lambda }_{km}}\hat{f}_{mk}^{*} \right|}^{2}}}+\sum\limits_{k'=1,k'\ne k}^{K}{p_{k'}}{{\left| \sum\limits_{m=1}^{M}{{{\lambda }_{km}}\hat{f}_{mk}^{*}{{f}_{mk'}}} \right|}^{2}}}$ in the real physical environment, or 
${\alpha}_{k}=\frac{{{p}_{k}}{{\left| \sum\limits_{m=1}^{M}{{{\lambda}_{km}} \hat{f}_{mk}^{*}{\hat{f}_{mk}}}\right|}^{2}}}{\sum\limits_{m=1}^{M}{{{\sigma_{m}^{2}\left| {{\lambda }_{km}}\hat{f}_{mk}^{*} \right|}^{2}}}+\sum\limits_{k'=1,k'\ne k}^{K}{p_{k'}}{{\left| \sum\limits_{m=1}^{M}{{{\lambda }_{km}}\hat{f}_{mk}^{*}{\hat{f}_{mk'}}} \right|}^{2}}}$ in the DT environment since the DT only records the estimated channels.
The achievable data rate of UE $k$ is $\mathcal{R}_k=\text{log}_2(1+ \alpha_k)$.

\section{Problem Formulation and Transformation}\label{section:problem_transformation}
In this section, we formulate a new problem that maximizes the achievable sum-rate of the considered system through interactions between the DT environment and learning algorithm. The maximization of the sum-rate is decoupled into two subproblems. A solution is developed for each subproblem.

\subsection{Problem Formulation}
The problem of interest is to optimize the AUA, power control, and RIS beamforming to maximize the sum-rate of the considered RIS-assisted, uplink UCCF system:
\begin{align*}
\mathcal P1: \quad \underset{\boldsymbol{\Lambda}, \mathbf p,\underline{\boldsymbol{\Theta}}} {\text{max}} &\sum\limits_{k=1}^{K}{{\mathcal R_k}} \nonumber\\
\text{s.t.} \ \, \text{C1}:\ \, &0\le {{p}_{k}}\le {p_{k,{\text{max}}}}, \ \, k \in\mathcal{K}, \forall m\in\mathcal{M},  \nonumber\\
\text{C2 :} \ \,  &\theta_{bn} \in \mathcal F_{\mathrm {c}}, \ \forall b\in\mathcal{B}, n\in\mathcal{N},\nonumber\\
\text{C3}: \ \,  &\sum_{m=1}^M \lambda_{km} \ge 1, \ \, k \in\mathcal{K}, \forall m\in\mathcal{M}, \nonumber\\
\text{C4}: \ \,  &\sum_{k=1}^K \lambda_{km} = 1, \ \, k \in\mathcal{K}, \forall m\in\mathcal{M}, \nonumber\\
\text{C5}: \ &\lambda_{km} \in \{0,1\}, \ \, k \in\mathcal{K}, \forall m\in\mathcal{M}, \nonumber\\
\text{C6}: \ \,  &\mathcal{R}_k \ge \mathcal{R}_{k,{\text{min}}}, \ \, \forall k \in\mathcal{K}, \nonumber
 \end{align*}
where $\mathbf p = [p_1, \cdots, p_K]^{\mathrm{T}}$ collects the transmit powers of all UEs; $p_{k,{\text{max}}}$ is the maximum transmit power of UE $k$; $\mathcal{R}_{k,{\text{min}}}$ is the required minimum data rate of UE $k$; $\underline{\boldsymbol{\Theta}}=\text{Reshape}(\boldsymbol{\theta},B,N)\in \mathbb{C}^{B\times N}$ is the reflection coefficient matrix of all RISs, also known as (a.k.a.) RIS beamforming matrix; and $\boldsymbol{\Lambda}$ is the AUA matrix with its $(k,m)$-th entry $\lambda _{km}$. 

Constraint C1 specifies the transmit power range  of UE $k$.
Constraint C2 specifies the phase shift of each reflection coefficient. C3 and C4 indicate that a UE can be served by multiple APs while an AP can only serve one UE at a time. Constraint C5 indicates the binary association decisions. Constraint C6 specifies the minimum transmit rate requirement $\mathcal{R}_{k,{\text{min}}}$ of UE $k$.

\subsection{Problem Transformation}
Problem $\mathcal{P}1$ is a mixed-integer nonlinear program (MINLP) due to its binary variable $\boldsymbol{\Lambda}$, continuous variables $\mathbf p$ and $\underline{\boldsymbol{\Theta}}$, and non-convex constraints C2 -- C6.
Traditional methods, such as fractional programming, convex optimization, and game theory, would incur a high computational complexity when solving problem $\mathcal{P}1$~\cite{Li2019Energy,ding2020deep,Wang2015VANET}.
Learning-based methods, such as DRL, can obtain effective solutions through interactions between the agent and the environment, and have demonstrated their effectiveness in solving complex problems~\cite{Li2019On,9473585}.

A conventional DRL-based design would learn all optimization variables (e.g., AUA, power control, and RIS beamforming) by a single DRL model, which would result in a complex action space comprising both discrete and continuous variables. The neural network would learn slowly and converge with difficulty.

To circumvent this impasse, we put forth a new learning framework, where
$\mathcal P1$ is decomposed into two iterative subtasks: The AUA subtask for discrete AUA decisions, and the PCRB subtask for two continuous decisions of power control and RIS beamforming. The resulting power control and RIS beamforming are collectively referred to as the PCRB result.  
By adopting the state-of-the-art TD3 model, the PCRB subtask is executed.
By developing a new PABPSO algorithm, the AUA subtask is efficiently accomplished.
An effective solution is achieved after several iterations of the AUA and PCRB subtasks.
Considering that a large number of interactions are required between the agents and environment in DRL models, we propose that the TD3 model interacts primarily with the virtual environment maintained by the DT and accesses the physical environment once for a while (or occasionally) to refine the learning. 

\section{Proposed DT-aided Learning Framework} \label{section:proposed_framework}
\begin{figure*}
	\centering{}\includegraphics[scale=0.89]{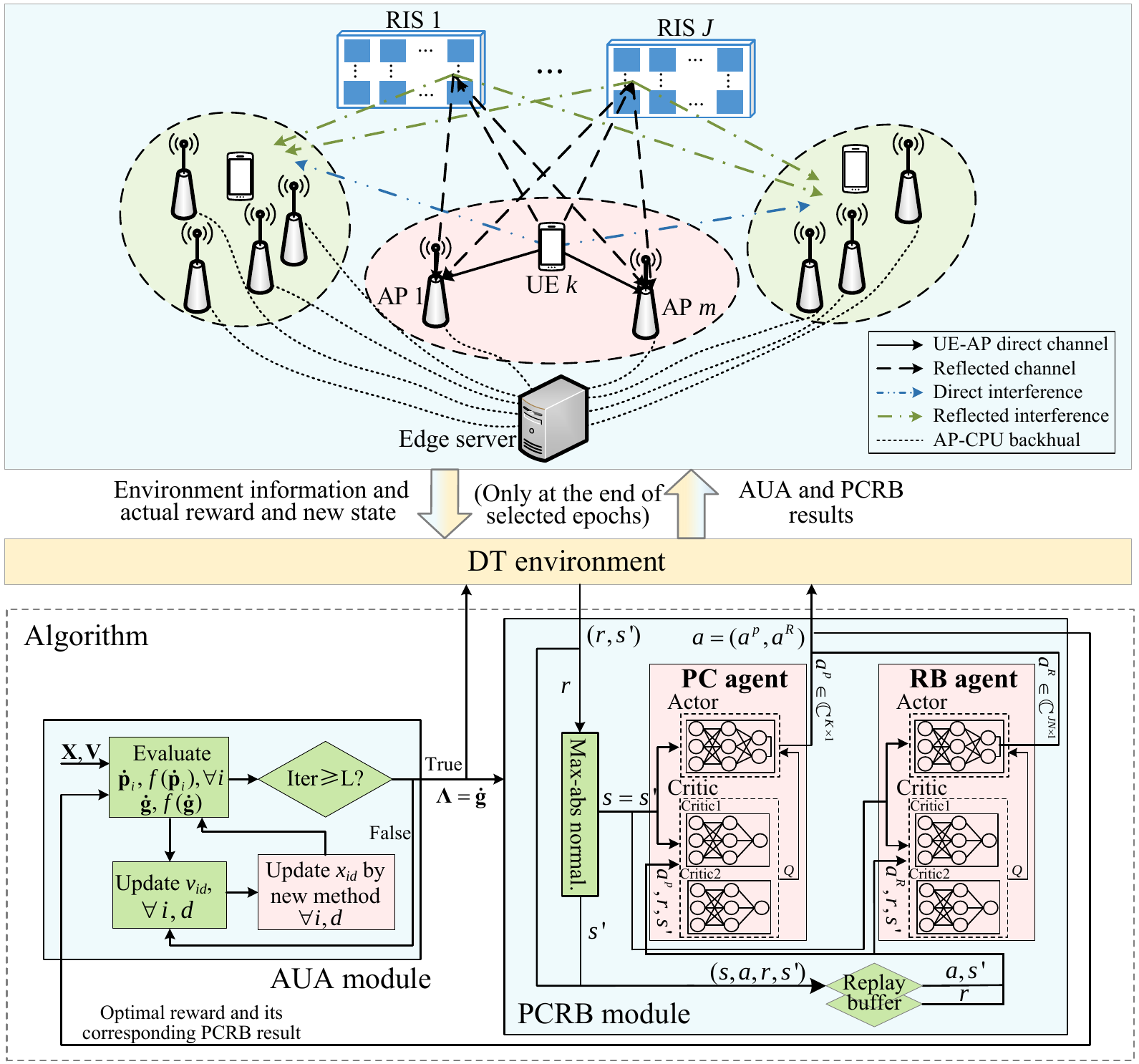}
	\caption{{\color{black} An illustration of the considered DT-aided learning for managing RIS-assisted UCCF systems, where  a DT is employed to serve as an intermediate layer between the physical environment and the proposed  framework. In the proposed framework, a new PABPSO-based AUA module is connected to a DRL-based PCRB module consisting of two parallel TD3 agents separately for power control and RIS beamforming.}}
	\label{fig:J_global}
\end{figure*}

In this section, we delineate the proposed learning framework that runs on the basis of epochs.
Each epoch starts with the AUA module that outputs $\boldsymbol{\Lambda}$ after $L$ iterations. $L$ is the pre-specified number of iterations for the convergence of the module. Given $\boldsymbol{\Lambda}$, the PC and RB agents of the PCRB module execute $Y$ episodes in parallel to obtain $\mathbf p$ and $\underline{\boldsymbol{\Theta}}$, respectively.
$Y$ is pre-configured, within which the PCRB module does not have to converge.
Let $Z$ denote the number of epochs needed for the convergence of the framework. The proposed framework is shown in Fig.~\ref{fig:J_global}.

During each epoch, each agent interacts only with the DT to test its learned policy and train its DNN. The DT serves as the virtual environment and returns the reward and new state to the agents. The agents may choose to interact with the physical environment at the end of some selected epochs to deploy or test their learned policies in the real world. 

\subsection{PABPSO-based AUA Module} \label{subsection:PABPSO}
We propose a new PABPSO method to solve the AUA subtask, which is an INLP problem with a finite discrete decision space.
Different from a typical BPSO method, the PABPSO method ensures that the positions of all particles satisfy Constraints C3 and C4, rather than only the binary Constraint C5, for $\boldsymbol{\Lambda}$ after each iteration, thereby leading to a substantially smaller search space and faster convergence than the conventional BPSO method \cite{637339}. Other existing methods, such as exhaustive search, relaxation \cite{Li2016Energy}, and discrete DRL algorithm \cite{9174775}, have been used to solve INLP problems. However, they are inapplicable to $\mathcal{P}1$. Exhaustive search suffers from exponentially growing solution spaces with increasing problem size, i.e., $\mathcal O(2^{KM})$.
Relaxation would reduce the accuracy of the AUA results and increases computational complexity \cite{9145209}.
Discrete DRL models, such as DQN and double DQN, would suffer from the curse of dimensionality. Particularly, there would be a total of $2^{KM}$ AUA decisions in problem $\mathcal P1$, implying $2^{KM}$ neurons in the output layer.

The new PABPSO algorithm improves the position updating, compared to the conventional BPSO method \cite{637339}.
A candidate solution to the AUA subproblem is interpreted as the position of a particle in a $D$-dimensional space, where $D=K M$ depends on the numbers of UEs and APs.
The fitness function of a particle is typically defined, according to the objective of the AUA.
Each iteration of the PABPSO
adjusts the particles' velocities based the local best and the global best solutions, updates their positions, and evaluates the fitness function of each position, followed by updating the best positions of itself and all particles.
Following are the details.

We first set a particle swarm with $I$ particles collected by the set $\mathcal{I}=\{1,\cdots,I\}$.
Let $\mathbf x_i$ denote the position of particle $i$, where $\mathbf x_i=[x_{i1},\cdots,x_{id},\cdots,x_{iD}]$ and the subscript ``$_{id}$'' ($d=1,\cdots,D$) indicates the $d$-th element of particle $i$. The positions of the particle swarm, collected by
$\mathbf X =\left[\mathbf x_{1}^{\mathrm T}, \cdots, \mathbf x_{I}^{\mathrm T} \right]^{\mathrm T}\in \mathbb R^{I\times D}$, provide the existing candidate solutions.
The velocities of the particle swarm, collected by $\mathbf V =\left[\mathbf v_{1}^{\mathrm T}, \cdots, \mathbf v_{I}^{\mathrm T} \right]^{\mathrm T}\in \mathbb R^{I\times D}$, give the probabilities with which an element of the particle’s solution will change to ``1'',
where $\mathbf v_i= [v_{i1},\cdots,v_{iD}]$.
Let $\mathbf {\underline X}_i \triangleq \text{Reshape}(\mathbf x_i,K,M)\in \mathbb{R}^{K\times M}$, and $\mathbf {\underline X}_i(k,m)$ denote the $(k,m)$-th element of $\mathbf {\underline X}_i$. By substituting $\mathbf {\underline X}_i(k,m)=\lambda_{km}$ into the SINR (see Section~\ref{section:systemmodel}),   
we obtain the fitness value of $\mathbf x_{i}$, denoted by $f(\mathbf x_{i})$, as given by
\begin{equation}
\label{eq:f_id}
	f(\mathbf x_{i}) = f(\mathbf {\underline X}_i)=\begin{cases}
	\sum\limits _{k=1}^{K}\mathcal R_{k}, & \text{if}\ K_{\mathrm {u}}=0;\\
	K_{\mathrm {u}} A_{\mathrm {u}}, & \text{if}\ K_{\mathrm {u}} \ne 0,
\end{cases}
\end{equation}
where $K_{\mathrm {u}}$ is the number of UEs that do not satisfy Constraint C6; and $A_{\mathrm {u}}$ is a negative constant.
A better position of a particle corresponds to a bigger $f(\mathbf x_{i})$ value.

At the $l$-th iteration of the AUA, $\mathbf{\dot p}_i^{(l)} \in \mathbb{R}^{1\times D}$ is the optimal position of particle $i$ found so far (i.e., a local best solution).  ${\dot p}_{id}^{(l)}$ is the $d$-th element of $\mathbf{\dot p}_i^{(l)}$. $\mathbf{\dot g}^{(l)}= \text{max}\{\mathbf{\dot p}_i^{(l)}, \forall i \in \mathcal I\}\in \mathbb{R}^{1\times D}$ is the optimal position of all particles found so far (i.e., the global best solution), i.e., ${\dot g}_{d}^{(l)}$ is the $d$-th element of $\mathbf{\dot g}^{(l)}$.
Given $\mathbf{\dot p}_i^{(l-1)}$ and $\mathbf{\dot g}^{(l-1)}$,
the velocity of the $d$-th element  is~\cite{9111671}
\begin{align}
\label{eq:BPSO_V}
{v}_{id}^{(l)}=&\omega \cdot {v}_{id}^{(l-1)}+{c}_{1} \cdot \mathcal{U}_1(0,1) \left({{\dot p}_{id}^{(l-1)}}-{x}_{{id}}^{(l-1)}\right)
+ {c}_{2} \cdot \mathcal{U}_2(0,1) \left({\dot g}_d^{(l-1)}-{x}_{{id}}^{(l-1)}\right),
\end{align}
where
$\omega \in (0,1)$ is the inertia factor, $\mathcal{U}_1(0,1)$ and $\mathcal{U}_2(0,1)$ produce two independent random uniform numbers between 0 and 1, and $c_1$ and $c_2$ are constant learning factors.

The position of particle $i$, $\mathbf x_i$, is updated based on its velocity $\mathbf v_i$ and Constraints C3 and C4 ($\mathbf x_i$ is a binary vector that inherently satisfies Constraint C5).
Let
$\mathbf {\underline V}_i^{(l)} \triangleq \text{Reshape}(\mathbf v_i^{(l)},K,M)\in \mathbb{R}^{K\times M}$, and $\mathbf {\underline V}_i^{(l)}(k,m)$ denote the $(k,m)$-th element of $\mathbf {\underline V}_i^{(l)}$.
$\mathbf {\underline X}_i^{(l)} \triangleq \text{Reshape}(\mathbf x_i^{(l)},K,M)$ is one of the potential solutions to the AUA.
A particle with a higher velocity has a higher probability of becoming ``1''.
We set an element to ``1'' in the $m$-th column of $\mathbf {\underline X}_i^{(l)}$, if the corresponding element in $\mathbf {\underline V}_i^{(l)}$ gives the highest velocity in that column of $\mathbf {\underline V}_i^{(l)}$.
The other elements in the $m$-th column of $\mathbf {\underline X}_i^{(l)}$ are set to ``0''. In other words,
$\mathbf {\underline X}_i^{(l)}(k,m)=1$,
if $\mathbf {\underline V}_i^{(l)}(k,m)=\max_k(\mathbf {\underline V}_i^{(l)}(k,m))$; or $\mathbf {\underline X}_i^{(l)}(k,m)=0$, otherwise.
As a result, there is only a single ``1'' element in each column of $\mathbf {\underline X}_i^{(l)}$, 
satisfying Constraint C4.

\begin{figure}[htp]
	\centering{}\includegraphics[scale=1.3]{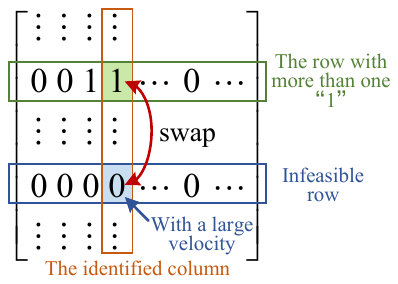}
	\caption{An example of the rectification of an infeasible row, where a ``1'' element is moved from a feasible row with more than one ``1'' element to an infeasible row.}
	\label{fig:BPSO_swi}
\end{figure}

Some UEs may undergo poor effective channels and are not served by any AP, and the corresponding rows of $\mathbf {\underline X}_i^{(l)}$ only have ``0'' elements. The rows violating Constraint C3 are referred to as \textit{infeasible rows}. 
We \textit{rectify the infeasible rows} one by one until all rows satisfy Constraint C3.
Since any feasible column of $\mathbf {\underline X}_i^{(l)}$ should have a single ``1'' element based on Constraint C4, an infeasible all-zero row is rectified by identifying a column and moving the ``1'' element of the column to the infeasible row without creating a new infeasible row; see Fig.~\ref{fig:BPSO_swi}. 

Preferably, the identified column corresponds to an element with a large velocity in the infeasible row,
to comply with the design criterion of particle position updating~\cite{9111671}. This is done by first checking the element with the highest velocity in an infeasible row of $\mathbf {\underline X}_i^{(l)}$. If the element can be updated by swapping its ``0'' element and the only ``1'' element in the same column without generating a new infeasible row, the current infeasible row is rectified, and we continue to rectify the next infeasible row. Otherwise, we proceed with the element with the second highest velocity in the infeasible row, and so on and so forth, until the infeasible row is rectified.
After all infeasible rows are rectified, $\mathbf {\underline X}_i^{(l)}$ satisfies both Constraints C3 and C4.

\subsection{DRL-based PCRB Module}
The PCRB subproblem is a non-convex optimization problem.
We design a new parallel DRL method consisting of two parallel agents deployed at the edge server: PC agent (for training the $K$ actions of power control) and RB agent (for training the $BN$ passive beamforming actions of the RIS elements), as shown in Fig.~\ref{fig:J_global}.
The PC agent and the RB agent have the same state space, reward function, and environmental information. Yet, they are trained separately and interactively. In contrast, the traditional DRL method using a single agent and producing all actions at the same time would be difficult to train because the transmit power and RIS beamforming have different value ranges and distributions and can lead to large errors in the gradient values.

In a state, the PC agent outputs the power control. The RB agent outputs the RIS beamforming.
The two agents interact with the environment (i.e., the DT during an epoch, and the physical environment via the DT at the end of some selected epochs) to obtain a reward for their current policy and transfer to the next state. The reward is evaluated based on the joint action of power control and RIS beamforming. The state, joint action, reward, and the next state are recorded as a tuple in a replay buffer shared by both agents.
By using batch learning \cite{xu2019targeted}, the PC and RB agents sample batches of tuples from the replay buffer to update their model parameters.

\subsubsection{State}
For the considered system, the environment state can be expressed as 
\begin{equation}\label{s_t}
\begin{aligned}
    \boldsymbol s_t^{\text{total}}=[
    \underbrace{\mathbf{\hat h}_{MK}^{\mathrm{T}},\mathbf{\hat h}_{MBNK}^{\mathrm{T}}
    }_{\boldsymbol s_t^{\mathrm {o}}},
    \underbrace{\boldsymbol {\alpha}_{t-1}^{\mathrm{T}},\mathbf{p}_{t-1}^{\mathrm{T}},\boldsymbol{\phi}_{t-1}^{\mathrm{T}}, \boldsymbol{\Lambda}^{\mathrm{T}}}_{\boldsymbol s_t^{\mathrm {d}}}],&
\end{aligned}
\end{equation}
where $\mathbf{\hat h}_{MK}=[\hat h_{11}, \cdots,\hat h_{MK}]^{\mathrm{T}} \in \mathbb C^{MK\times 1}$ and $\mathbf{\hat h}_{MBNK}=[\hat h_{1111},\cdots,
\hat h_{MBNK}]^{\mathrm{T}}\in \mathbb C^{MBNK\times 1}$ collect the
estimated channels, 
and $\boldsymbol{\Lambda}=[\lambda_{11},\cdots,\lambda_{km}]^{\mathrm T} \in \mathbb R^{MK\times 1}$.
The size of $\boldsymbol s_t^{\text{total}}$ is 
$(3MK+2MBNK+2K+BN)$.

We note that $\boldsymbol s_t^{\text{total}}$ can be divided between an infrequently changing part $\boldsymbol s_t^{\mathrm {o}}$ and a frequently changing part $\boldsymbol s_t^{\mathrm {d}}$. The infrequently changing part, $\boldsymbol s_t^{\mathrm {o}}$, is input to evaluate the reward accurately only once after the DRL interacts with the physical environment at the end of some selected epochs. The frequently changing part of the environment state, $\boldsymbol s_t^{\mathrm {d}}$, can be further divided into two parts, namely, epoch state information and step state information.  

The epoch state information, $\boldsymbol{\Lambda}$, does not change during an epoch and changes between epochs. It is input into the PCRB module once per epoch to evaluate the reward of the PCRB.
The step state information, $\boldsymbol{\alpha}_{t-1},\mathbf{p}_{t-1}$ and $\boldsymbol{\phi}_{t-1}$, changes between steps. 
Hence, this part of the state needs to be observed by the PC and RB agents at each step $t$, and can be written as 
\begin{align}
\label{eq:s_t}
    \boldsymbol s_t=[\boldsymbol {\alpha}_{t-1}^{\mathrm{T}},\mathbf{p}_{t-1}^{\mathrm{T}},\boldsymbol{\phi}_{t-1}^{\mathrm{T}}]. 
\end{align}
The size of $\boldsymbol s_t$ is only $(2K+BN)$.

\subsubsection{Action}
As for the PC agent, the action is $\boldsymbol a_t^{\mathrm p}=[a_{1,t}^{\mathrm p},\cdots,a_{K,t}^{\mathrm p}]^{\mathrm{T}}$, where $-1 \leq a_{k,t}^{\mathrm p} \leq 1$, $\forall k$.
As for the RB agent, we take the phase shifts of all the RISs, $\phi_{bn}$, as the output, since the RIS beamforming $\theta_{bn}=e^{\jmath\phi_{bn}}$ has a unit-modulus and complex value but existing DNN implementations do not support complex outputs. The action $\boldsymbol{a}_t^{\mathrm R}$ is $\boldsymbol{a}_t^{\mathrm R}=[a_{1,t}^{\mathrm R},\cdots,a_{BN,t}^{\mathrm R}]^{\mathrm{T}} \in \mathbb R^{BN\times 1}$, where $-1 \leq a_{bn,t}^{\mathrm R} \leq 1$, $\forall b,n$. The action space $\mathcal A$ is the pair of actions $\boldsymbol a =(\boldsymbol a^{\mathrm p},\boldsymbol a^{\mathrm R})$.

The transmit powers of the UEs and the phase shift of the RISs are used to calculate the reward, where
$p_{k,t} \triangleq \frac{(a_{k,t}^{\mathrm p}+1)}{2} p_{k,\text{max}}$ is the transmit power of UE $k$, and $\mathbf{p}_t=[p_{1,t},\cdots, p_{K,t}]^{\mathrm{T}}$. $\phi_{bn,t}=(a_{bn,t}^{\mathrm R}+1)\pi$ is the phase shift of the $n$-th element of RIS $b$, and $\boldsymbol{\phi}_t = [\phi_{11,t},\cdots, \phi_{BN,t}]^{\mathrm{T}}$.

\subsubsection{Reward}
The immediate reward of the PCRB module consists of the system sum-rate and a penalty, accounting for the optimization objective and constraints, respectively.
{\color{black}At step $t$, the immediate reward (returned by the DT during an epoch, or returned by the physical environment at the end of a selected epoch) is defined as}
\begin{equation}
\label{eq:reward}
	r_t = \begin{cases}
	\sum\limits_{k=1}^{K}\mathcal R_{k,t}, & \text{if}\ K_{\mathrm c} =0,\\
	K_{\mathrm c} A_{\mathrm {c}}, & \text{if}\ K_{\mathrm c} \ne 0,
\end{cases}
\end{equation}
where $K_{\mathrm c}$ is the number of infeasible action pairs violating Constraint C6, and $A_{\mathrm {c}}$ is the penalty factor. $A_{\mathrm {c}} < 0$ depends on $\textstyle\sum_{k=1}^{K}\mathcal R_{k,t}$.
To ensure stable convergence of each agent, $\left|K_{\mathrm c} A_{\mathrm {c}}\right|$ and $\sum_{k=1}^{K}\mathcal R_{k,t}$ need to be in the same order of magnitude. If $\left|K_{\mathrm c} A_{\mathrm {c}}\right|$ is too large, the positive reward (i.e., $\sum_{k=1}^{K}\mathcal R_{k,t}$) is too low to motivate the agent. On the other hand, if $\left|K_{\mathrm c} A_{\mathrm {c}}\right|$ is too small, the actions of the agents may not meet the constraints. 

\subsubsection{Update algorithm of TD3}
Let $\tau$ and $T$ denote the number of steps executed since step $t$ and the total step number of the PCRB module, respectively. $\gamma \in [0,1)$ is the discount factor.
The action-value function (Q-function) defines the expected reward for action $\boldsymbol a$ taken following policy $\pi_{\boldsymbol s}$ in the state $\boldsymbol s$, and satisfies the following Bellman equation \cite{sutton2018reinforcement}
\begin{align}
\label{eq:Q_F}
Q^{\pi}(\boldsymbol s, \boldsymbol a)&=\mathbb{E}_{\pi_{\boldsymbol s}}\left\{\sum_{\tau=t}^{T} \gamma^{\tau-t} r_{\tau} \mid \boldsymbol s_{t}=\boldsymbol s, \boldsymbol a_{t}=\boldsymbol a\right\} \\ \nonumber
&=\mathbb{E}_{\pi_{\boldsymbol s}}\{r_{t}+\gamma Q^{\pi}\left(\boldsymbol s_{t+1}, \boldsymbol a_{t+1}\right) \mid \boldsymbol s_{t}=\boldsymbol s, \boldsymbol a_{t}=\boldsymbol a\}.
\end{align}

DDPG \cite{2015Continuous} is a classical algorithm for solving continuous state and action spaces using an actor-critic (AC) framework. It combines the value-based method (critic) and the policy-based method (actor).
Let $\mu$ denote the parameter of the policy network and $\iota(\cdot)$ denote the mapping function of the neural network. DDPG solves $\max _{\boldsymbol a} Q(\boldsymbol s, \boldsymbol a) \approx Q(\boldsymbol s, \iota(\boldsymbol s \mid \mu))$ to find the action output that maximizes the Q value.
An overestimation bias is an observed problem of DDPG. The accumulated error can result in unreasonable states estimated with high values, leading to suboptimal policy updates and divergent behaviors \cite{fujimoto2018addressing}.

The state-of-the-art TD3 \cite{fujimoto2018addressing} can alleviate the overestimation problem in DDPG. The general TD3 has three key improvements compared to the DDPG: 1) Using two sets of critic-networks (twin) to output two Q-values, namely, clipped double Q-learning, thereby suppressing continuous overestimated biases; 2) Delayed policy updates: The actor is updated after the critic has been updated multiple times, hence reducing the accumulated error and the variance of the approximate Q-function; and 3) Target policy smoothing: Adding random noises to the target action to improve the robustness of the Q-target valuation.

The standard TD3 model is not directly suitable for the PCRB subtask through, since the state $\boldsymbol s_t$ consists of the SINRs, transmit powers, and phase shifts, the three of which can have substantially different ranges and distributions. The difference can compromise the convergence speed, and the effectiveness of the neural network activation function.
We add a pre-processing module, max-abs normalization \cite{articlePrognosisModel}, to the TD3 model to handle the state data without destroying the original data distribution.

As shown in Fig.~\ref{fig:J_global}, we use two TD3 modules in parallel for the PC and RB agents.
A TD3 model includes six networks, i.e., two evaluated-critic-networks, two target-critic-networks, an evaluated-actor-network, and a target-actor-network. Let $\zeta_i$ denote the parameters of the $i$-th evaluated-critic-network with the output $Q_{\zeta_i}$, and $\zeta_i^{\prime}$ denote the parameters of the $i$-th target-critic-network with the output $Q_{\zeta_i^{\prime}}$, $i=1,2$.
$\mu$ and $\mu^{\prime}$ denote the parameters of the evaluated-actor-network and target-actor-network with the outputs $\boldsymbol a$ and $\boldsymbol {\tilde{a}}$, respectively, $\boldsymbol a \in[\boldsymbol a_{\text{min}}, \boldsymbol a_{\text{max}}]$. 
Under policy $\pi_{\mu^{\prime}}$ with parameters ${\mu^{\prime}}$ and the target state $\boldsymbol s^{\prime}$, the target action
is $\boldsymbol {\tilde{a}} \leftarrow\text{clip}(\pi_{\mu^{\prime}}(\boldsymbol s^{\prime})+\epsilon,\boldsymbol a_{\text{min}},\boldsymbol a_{\text{max}})$, where $\epsilon \sim \text{clip} (\mathcal{N}(0, \tilde{\sigma}^2),-c , c)$. The target action noise is clipped to be within the range $[-c, c]$. The target update of the clipped double Q-learning follows
$y=r+\gamma\mathrm{min}_{i=1,2} Q_{\zeta_{i}^{\prime}}(\boldsymbol s^{\prime}, \boldsymbol {\tilde{{a}}})$.

The inputs of the TD3 are batches randomly taken from the experience replay buffer, after which the TD3 updates the weights of the six neural networks. $\zeta_i$, $i=1,2$ is updated by
\begin{equation}
\zeta_{i} \leftarrow \text{argmin}_{\zeta_{i}} C_{\mathrm {B}}^{-1} \sum\left(y-Q_{\zeta_{i}}(\boldsymbol s, \boldsymbol a)\right)^{2},
\end{equation}
where $C_{\mathrm {B}}$ is the batch size. The actor-network is resistant to the overestimation of the $Q$ value. It searches for the maximum $Q$ value by running the gradient ascent. 

As the update proceeds, $Q_{\zeta_{1}}$ and $Q_{\zeta_{2}}$ would become increasingly conformed. The actor can choose either $Q_{\zeta_{1}}$ or $Q_{\zeta_{2}}$ to update $\mu$, as given by
\begin{equation}
\nabla_{\mu} J(\mu)=\left.B^{-1} \sum \nabla_{\boldsymbol a} Q_{\zeta_{1}}(\boldsymbol s, \boldsymbol a)\right|_{\boldsymbol a=\pi_{\mu}(\boldsymbol s)} \nabla_{\mu} \pi_{\mu}(\boldsymbol s).
\label{eq:TDpolicy}
\end{equation}
Then, $\zeta_{i}^{\prime}$, $i=1,2$, and $\mu^{\prime}$ are updated by soft update:
\begin{align}
\zeta_{i}^{\prime} \leftarrow \chi \zeta_{i}+(1-\chi) \zeta_{i}^{\prime};\, \mu^{\prime} \leftarrow \chi \mu+(1-\chi) \mu^{\prime},
\label{eq:ruangengxin}
\end{align}
where $\chi \in (0,1)$ is the coefficient of the soft update.

\subsection{Training Algorithm}
The input of the AUA module includes the power control and the phase shift decisions by the PCRB module, denoted respectively by $\mathbf{p}^{\text{AUA}}$ and $\boldsymbol{\phi}^{\text{AUA}}$, 
$\mathbf{\dot p}_i$, $f(\mathbf{\dot p}_i)$, $\mathbf{\dot g}$, $f(\mathbf{\dot g})$ and the environment information. The output of the AUA module is the AUA, $\boldsymbol{\Lambda}$.
The overall training process is summarized in Algorithm \ref{alg:A1}, and the training process of the PCRB module is summarized in Algorithm \ref{alg:PCRB}.
Here, the superscripts ``$^\mathrm{p}$'' and ``$^\mathrm{p'}$'' indicate the evaluated- and  target-networks of the PC agent, respectively. The superscripts ``$^\mathrm{R}$'' and ``$^\mathrm{R'}$'' indicate the evaluated- and target-networks of the RB agent, respectively.

Note that initialization is important for the convergence of the AUA module. We design a \emph{Merit-based input} mechanism, where the optimal reward $r^{\text{opt}}$ per epoch of the PCRB module and the corresponding optimal action $(\mathbf{p}^{\text{opt}}$
$\boldsymbol{\phi}^{\text{opt}})$, are input to the AUA module.
Specifically, we initialize $f(\mathbf{\dot g})$, $\mathbf{p^{\text{AUA}}}$, and $\boldsymbol{\phi}^{\text{AUA}}$ at epoch $z$ with $r^{\text{opt}}$, $\mathbf{p}^{\text{opt}}$ , and $\mathbf{\phi}^{\text{opt}}$ at epoch $z-1$, respectively.

\begin{algorithm}
\small
\caption{New learning-based framework for maximizing the sum-rate of the new RIS-assisted, uplink, UCCF systems.}
\label{alg:A1}
\begin{algorithmic}[1]
\REQUIRE
\STATE {Initialize the PC agent and the RB agent, $\zeta_{1}^{\mathrm {p}}$, $\zeta_{2}^{\mathrm {p}}$, $\mu^{\mathrm {p}}$, $\zeta_{1}^{\mathrm {R}}$, $\zeta_{2}^{\mathrm {R}}$ and $\mu^{\mathrm {R}}$; and assign $\zeta_{1}^{\mathrm {p'}} \leftarrow \zeta_{1}^{\mathrm {p}}$, $\zeta_{2}^{\mathrm {p'}} \leftarrow \zeta_{2}^{\mathrm {p}}$,  $\mu^{\mathrm {p'}} \leftarrow \mu^{\mathrm {p}}$,
$\zeta_{1}^{\mathrm {R'}} \leftarrow \zeta_{1}^{\mathrm {R}}$, $\zeta_{2}^{\mathrm {R'}} \leftarrow \zeta_{2}^{\mathrm {R}}$,
$\mu^{\mathrm {R'}} \leftarrow \mu^{\mathrm {R}}$;}
\STATE {Initialize the experience replay buffer, $\mathbf{p}^{\text{AUA}}$, $\boldsymbol{\phi}^{\text{AUA}}$, and $r^{\text{opt}}$; and input $\boldsymbol{s}_t^{\mathrm {o}}$ to DT;}
\STATE{Initialize $\mathbf{\dot g}$ and $f(\mathbf{\dot g})$ of PABPSO;}
\FOR{epoch $z=1, \cdots , Z$}
\STATE{Initialize a particle swarm with $\mathbf X$, $\mathbf V$, and initialize $\mathbf{\dot p}_i$ and $f(\mathbf{\dot p}_i)$ of each particle $i$, $\forall i$;}\\
{\vspace{2 mm}\underline{\textbf{PABPSO-based AUA:}}}\nonumber\\
\FOR{iter $l=1, \cdots, L$}
\FOR{particle $i=1,\cdots,I$}
\STATE {Update ${v}_{id}^{(l)}\,\forall d$ by (\ref{eq:BPSO_V});}\\
Update ${x}_{id}^{(l)}$ $\forall d$ by
setting $\mathbf {\underline X}_i^{(l)}(k,m)\!=\!1$, if $\mathbf {\underline V}_i^{(l)}(k,m)\!=\!\max_k(\mathbf {\underline V}_i^{(l)}(k,m))$, or $\mathbf {\underline X}_i^{(l)}(k,m)\!=0$, otherwise, for $m=1,\cdots,M$; \\
\STATE{Rectify any infeasible rows, see Section~\ref{subsection:PABPSO};}
\STATE{\textbf{If} $f(\mathbf x_{i})>f(\mathbf{\dot p}_i)$, \textbf{then} $\mathbf{\dot p}_i=\mathbf x_{i}$; 
}
\STATE{\textbf{If} $f(\mathbf{\dot p}_i)>f(\mathbf{\dot g})$, \textbf{then} $\mathbf{\dot g}=\mathbf{\dot p}_i$;
}
\ENDFOR
\ENDFOR\\
{\vspace{2 mm}\underline{\textbf{TD3-based PCRB:}}}\nonumber\\
\FOR{episode $y=1, \cdots, Y$}
\STATE {Initialize the DT environment based on $\boldsymbol{\Lambda}\!=\!{\text{Reshape}}(\mathbf{\dot g},K,M)$, and observe the initial state;}
\FOR{step $t=1, \cdots, T$}
\STATE{Input $\boldsymbol s$ to the PC and RB agents;}
\STATE{Output
$\boldsymbol a=(\boldsymbol a^{\mathrm p},\boldsymbol a^{\mathrm R})$, and convert 
to the action $(\mathbf p,\boldsymbol{\phi})$;}
\STATE{Execute $\boldsymbol a$, and observe $r$ and the new state $\boldsymbol s'$;}
\STATE{\textbf{If} $r>r^{\text{opt}}$, \textbf{then} $r^{\text{opt}}=r$, $\mathbf p^{\text{opt}}=\mathbf p$ and $\boldsymbol{\phi}^{\text{opt}}=\boldsymbol{\phi}$;
}
\STATE{Store ($\boldsymbol s$, $\boldsymbol a$, $r$, $\boldsymbol s'$) in the replay buffer;}
\STATE{Run \textbf{Algorithm \ref{alg:PCRB}};}
\ENDFOR
\ENDFOR
\IF{the DRL interacts with the physical environment}
\STATE{DRL observes the real reward and the real next state;}
\ENDIF
\STATE{Update $\mathbf{p}^{\text{AUA}}=\mathbf p^{\text{opt}}$, $\boldsymbol{\phi}^{\text{AUA}}=\boldsymbol{\phi}^{\text{opt}}$, and $f(\mathbf{\dot g})=r^{\text{opt}}$.}
\ENDFOR
\end{algorithmic}
\end{algorithm}

\begin{algorithm}[htp]
\small
\caption{Training process of the PC and RB agents}
\label{alg:PCRB}
\begin{algorithmic}[1]
\STATE{Sample a mini-batch of $C_{\mathrm {B}}$ transitions $(\boldsymbol s,\boldsymbol a, r, \boldsymbol s')$ from the experience replay buffer;}
\STATE{Obtain $\boldsymbol a^{\mathrm p}$ and $\boldsymbol a^{\mathrm R}$  from $\boldsymbol a$;}\\
\STATE{Add truncated noise and obtain $\tilde{\boldsymbol a}^{\mathrm p}(\mathbf{s'})$ by $\tilde{\boldsymbol a}^{\mathrm p}(\mathbf{s'})=\text{clip}(\pi_{{\mu^{\mathrm {p'}}}}(\mathbf{s'})+\epsilon^{\mathrm p}, \boldsymbol a_{\text{min}}^{\mathrm p}, \boldsymbol a_{\text{max}}^{\mathrm p})$,\\ $\epsilon^{\mathrm p} \sim \text{clip} (\mathcal{N}(0, \tilde{\sigma^{\mathrm p}}^2),-c^{\mathrm p} , c^{\mathrm p})$;}
\STATE{Double Q-network for target policy: $y=r+ \gamma \underset{i=1, 2 }{\text{min}}Q_{\zeta_i^{\mathrm {p}}}(\mathbf{s'},\tilde{\boldsymbol a}^{\mathrm p})$;}
\STATE{Update $\zeta_i^{\mathrm {p}}\leftarrow \text{min}_{\zeta_i^{\mathrm {p}}}\frac{1}{C_{\mathrm {B}}}\sum(y-Q_{\zeta_i^{\mathrm {p}}}(\boldsymbol s,\boldsymbol a^{\mathrm p}))^2$;}
\STATE{For the RB agent,
$\tilde{\boldsymbol a}^{\mathrm R}(\mathbf{s'})=\text{clip}(\pi_{{\mu^{\mathrm {R'}}}}(\mathbf{s'})+\epsilon^{\mathrm R}, \boldsymbol a_{\text{min}}^{\mathrm R}, \boldsymbol a_{\text{max}}^{\mathrm R})$,
$\epsilon^{\mathrm R} \sim \text{clip} (\mathcal{N}(0, \tilde{\sigma^{\mathrm R}}^2),-c^{\mathrm R} , c^{\mathrm R})$,\\ $y=r+ \gamma \underset{i=1, 2 }{\text{min}}Q_{\zeta_i^{\mathrm {R}}}(\mathbf{s'},\tilde{\boldsymbol a}^{\mathrm R})$;}
\STATE{Update $\zeta_i^{\mathrm {R}}\leftarrow \text{min}_{\zeta_i^{\mathrm {R}}}\frac{1}{C_{\mathrm {B}}}\sum(y-Q_{\zeta_i^{\mathrm {R}}}(\boldsymbol s,\boldsymbol a^{\mathrm R}))^2$;}
\IF{$t$ mod $T_\mathrm{A}$ = 0}
\STATE{Update ${\mu^{\mathrm p}}$ by
$\nabla_{\mu^{\mathrm p}} J(\mu^{\mathrm p}) = \frac{1}{C_{\mathrm {B}}} \sum \nabla_{\boldsymbol a^{\mathrm p}}Q_{\mu_1^{\mathrm p}}(\boldsymbol s,  \boldsymbol a^{\mathrm p})|_{\boldsymbol a^{\mathrm p} = \pi_{\mu^{\mathrm p}}(\boldsymbol s)} \nabla_{\mu^{\mathrm p}} \pi_{\mu^{\mathrm p}}(\boldsymbol s)$;}
\STATE{Update $\mu_i^{\mathrm p'}$ ($i=1,2$) and $\mu^{\mathrm p'}$ by
$\mu_i^{\mathrm p'} \leftarrow \chi \mu_i^{\mathrm p}+(1-\chi) \mu_i^{\mathrm p'};\, \mu^{\mathrm p'} \leftarrow \chi \mu^{\mathrm p} +(1-\chi) \mu^{\mathrm p'}$;}
\STATE{Update ${\mu^{\mathrm R}}$ by 
$\nabla_{\mu^{\mathrm R}} J(\mu^{\mathrm R}) = \frac{1}{C_{\mathrm {B}}} \sum \nabla_{\boldsymbol a^{\mathrm R}} Q_{\mu_1^{\mathrm R}} ( \boldsymbol s, \boldsymbol a^{\mathrm R} ) |_{\boldsymbol a^{\mathrm R} = \pi_{\mu^{\mathrm R}}(\boldsymbol s)} \nabla_{\mu^{\mathrm R}} \pi_{\mu^{\mathrm R}} (\boldsymbol s)$;}
\STATE{Update $\mu_i^{\mathrm R'}$ ($i=1,2$) and $\mu^{\mathrm R'}$ by
$\mu_i^{\mathrm R'} \leftarrow \chi \mu_i^{\mathrm R}+(1-\chi) \mu_i^{\mathrm R'};\, \mu^{\mathrm {R'}} \leftarrow \chi \mu^{\mathrm R} +(1-\chi) \mu^{\mathrm {R'}}$;}
\ENDIF
\end{algorithmic}
\end{algorithm}

The time-complexity of the AUA module is $\mathcal{O}(ZLIKM)$, and it depends on the size of the particle swarm and the number of iterations.
The time-complexity of a DNN depends on its size~\cite{9575181}. Hence, the complexity of the PC agent is $\mathcal{O}(ZYT[(3K+BN)n_{\mathrm {pa}} +(3K+BN+1)(n_{\mathrm {pc1}}+n_{\mathrm {pc2}})+ (N_{\mathrm {pa}}-1)n_{\mathrm {pa}}^2+(N_{\mathrm {pc1}}-1)n_{\mathrm {pc1}}^2+(N_{\mathrm {pc2}}-1)n_{\mathrm {pc2}}^2])$.
The complexity of the RB agent is $\mathcal{O}(ZYT[(2K+2BN)n_{\mathrm{ra}} +(2K+2BN+1)(n_{\mathrm{rc}1}+n_{\mathrm{rc}2})+(N_{\mathrm{ra}}-1)n_{\mathrm{ra}}^2+(N_{\mathrm{rc}1}-1)n_{\mathrm{rc}1}^2+(N_{\mathrm{rc}2}-1)n_{\mathrm{rc}2}^2])$.
This is because the PC agent and the RB agent consist of DNNs and execute $ZYT$ steps.
Here, $n_{\mathrm {pa}}$, $n_{\mathrm {pc1}}$, and $n_{\mathrm {pc2}}$ are the numbers of neurons in the hidden layer of the actor-network, the first and second critic-networks in the PC agent, respectively. $N_{\mathrm {pa}}$, $N_{\mathrm {pc1}}$ and $N_{\mathrm {pc2}}$ are the numbers of hidden layers of the actor-network, the first and second critic-network in the PC agent, respectively. The subscripts ``$_{\mathrm{ra}}$'', ``$_{\mathrm{rc1}}$'', and ``$_{\mathrm{rc2}}$'' indicate the actor-network, the first and second critic-networks in the RB agent, respectively.
The time-complexity of the PCRB module is the larger of those of the PC and RB agents since the two agents run in parallel.
The total time-complexity of the proposed method captures both the AUA and PCRB modules.

The space-complexity of the proposed method, measuring the memory consumed by the algorithm, adds up those of the AUA module, PCRB module, and replay buffer.
The space-complexity of the AUA module is $\mathcal{O}(ZLIKM)$.
The space-complexity of the PCRB module is the sum of the complexity of the PC agent and RB agent, as discussed above.
Let $C_{\mathrm {r}}$ denote the capacity of the replay buffer. The space-complexity of the replay buffer is
$\mathcal{O}(C_{\mathrm {r}} \times(5K+3BN+1))$, as it depends on the capacity of the buffer and the size of the stored tuples.

\section{NUMERICAL RESULTS AND DISCUSSIONS}\label{section:sim}

In this section, extensive simulations are carried out to assess the effectiveness of the proposed algorithm and the value of the DT in terms of overhead reduction.

\subsection{Simulation Settings and Benchmarks}

Suppose that the UEs and APs are randomly uniformly distributed in a circular region with a radius of 100 m. The RISs are uniformly located along the 45$^{\circ}$ diagonal within the circle. 
The heights of the APs, UEs, and RISs are $h_{\text{AP}}= 15$ m, $h_{\text{UE}}= 1.5$ m, and $h_{\mathrm {RIS}}= 20$ m. The carrier frequency is $f_{\mathrm {c}}=$ 1.9 GHz. The signal bandwidth is 20 MHz.
The power spectral density of the noise is $\sigma_{m}^2= -106$ dBm/Hz at each AP $m$.
The standard deviation of shadowing is $\sigma_{\text{sh}}=$ 8 dB. The channel coherence time is $\tau_{\mathrm {c}}=$ 200 symbols. The uplink pilot lasts 8 symbols.  
All users have the same maximum transmit powers and minimum rate requirements, i.e., $p_{k,{\text{max}}}=p_{\text{max}},\, \mathcal R_{k,{\text{min}}}=\mathcal R_{\text{min}}$, $\forall k$.
The path losses follow the three-slope COST Hata model~\cite[eqs. (32) \& (33)]{9322151}.

The number of epochs is $Z= 18$.
For the AUA module, we set $I=$ 200, $c_1=c_2=$ 2, $v_{id}=[-10,10]$, $w=0.5$, $L=$ 80, and $A_{\mathrm {u}}= -3$.
For the PCRB module, we set $Y=$ 200, $T=$ 200, $C_{\mathrm {B}} =$ 128, $C_{\mathrm {r}}= 40,000$, $A_{\mathrm {c}}= -3$, and $\chi=0.001$.
We also set the exploration probability to 0.9 for both the PC  and  RB agents, and update $\mu^{\mathrm {p'}}$ and $\mu^{\mathrm {R'}}$ every $T_\mathrm{A}=$ 5 steps, where the Polyak averaging hyper-parameter is $\tau_{\mathrm{a}}=$ 0.001.
The noises for smoothing the target-actor-networks of the PC and RB agents are $(\tilde{\sigma^{\mathrm p}}^2, c^{\mathrm p})=(\frac{1}{2}a^{\mathrm p}_{\text{max}},a^{\mathrm p}_{\text{max}})$ and $(\tilde{\sigma^{\mathrm R}}^2,c^{\mathrm R})=(\frac{1}{2}a^{\mathrm R}_{\text{max}},a^{\mathrm R}_{\text{max}})$, respectively.
The activation functions in all hidden layers are rectified linear units (ReLU), and a final $\tanh(\cdot)$ unit following the output of the actor-networks.
The critic-networks are updated after the actor-networks are updated three times.
The discount factor is $\gamma=0.997$. The critic optimizer is Adam with  default hyper-parameters $\beta_1=$ 0.9 and $\beta_2=$ 0.999.

{\color{black}For comparison purposes, we also implement 
PC\&RB out~\cite{9110869}, Allout~\cite{9345106}, PABPSO-DDPG~\cite{2015Continuous}, and BPSO-TD3~\cite{9111671}.
In the PC\&RB out algorithm, both power control and RIS beamforming decisions are output by a TD3 model, and the AUA is output by PABPSO.
In the Allout algorithm, the AUA,  power control, and RIS beamforming decisions are all generated by a TD3 model.
The PABPSO-DDPG algorithm~\cite{2015Continuous} replaces the two TD3 models with two DDPG models in the proposed framework.
The BPSO-TD3 algorithm replaces the PABPSO with the standard BPSO in the proposed framework.
}
\subsection{Result and Discussion}
\begin{figure}
	\centering{}\includegraphics[scale=0.29]{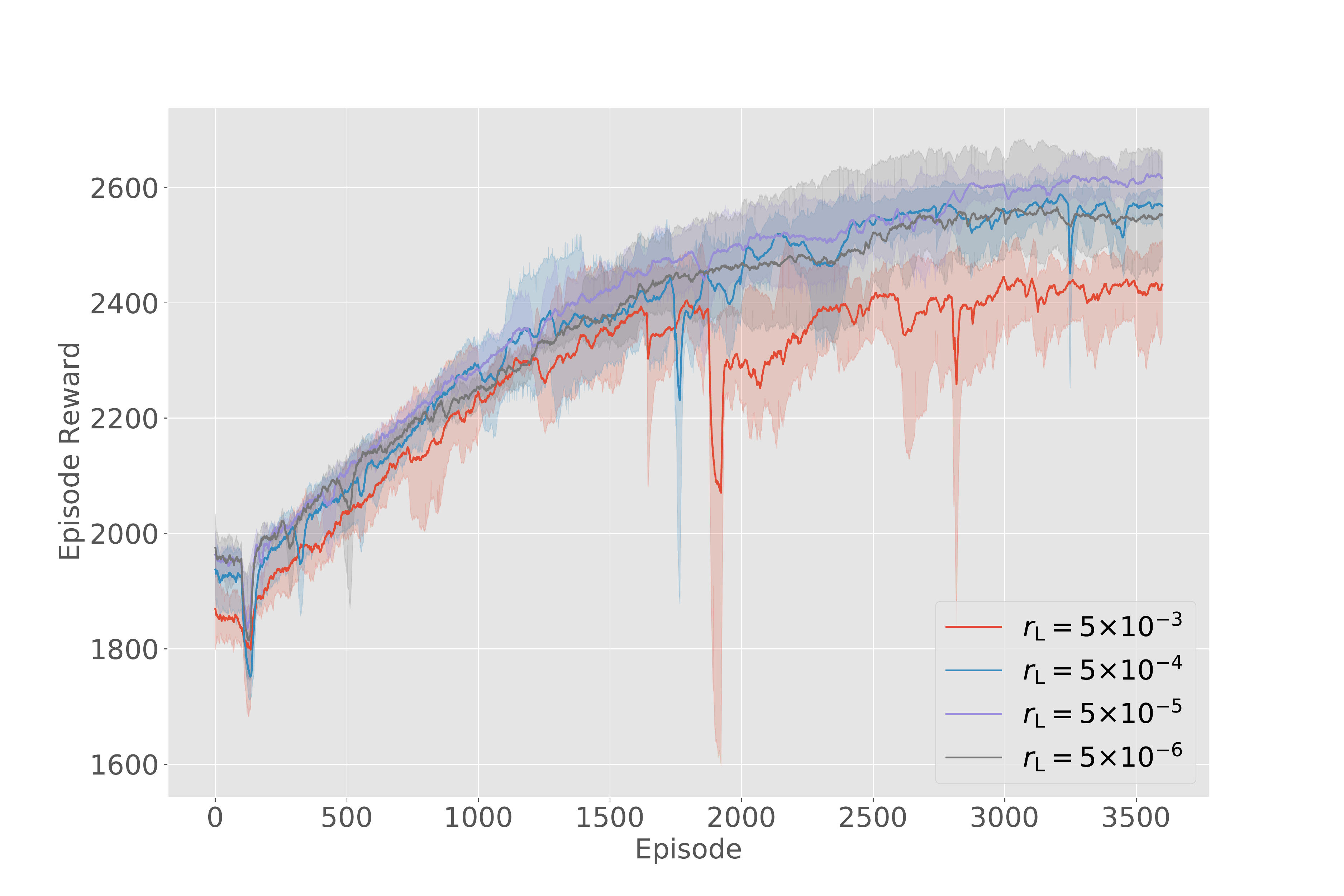}
	\caption{{\color{black}Convergence curves of episode rewards with different learning rates $r_{\mathrm L}$. ($K=$ 3, $M=$ 4, $B=$ 2, $N=$ 5, $p_{\text{max}}=$ 0.4 W and $\mathcal R_{\text{min}}=$ 0.2 bit/s.)}}
	\label{fig:learn}
\end{figure}

{\color{black}
Fig.~\ref{fig:Conver_curve} demonstrates the convergence of the proposed algorithm by plotting the episode reward with the increase of episodes, where the episode reward is the accumulated step reward in an episode of the PCRB module.
We compare the different settings where the proposed algorithm can test its model in the physical environment at the end of each epoch or at the end of every two epochs, or does not test the model in the physical environment until the end of training. We also plot the ideal case where the channel estimations are error-free. The ideal case can be interpreted as a situation where no DT is employed, and the proposed DRL model directly interacts with the physical environment once per step.
It is observed that the proposed algorithm converges under all considered settings.
\begin{figure}[t]
\centering
\begin{minipage}[t]{0.48\textwidth}
\centering
\includegraphics[width=7.9cm]{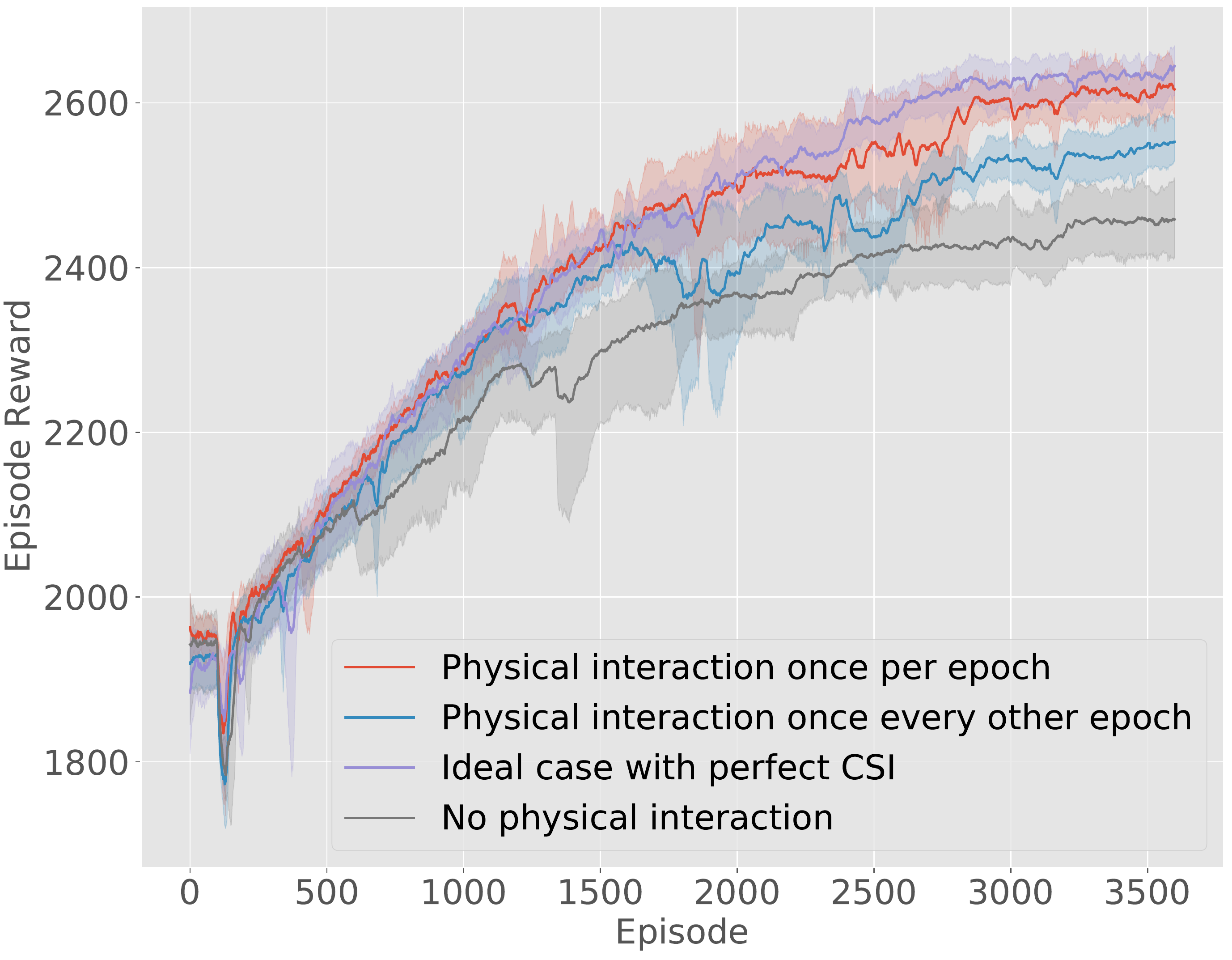}
\caption{The convergence curves of episode rewards over episodes under different interaction frequencies between the DT and physical environment. $K=$ 3, $M=$ 4, $B=$ 2, $N=$ 5, $p_{\text{max}}=$ 0.4 W, and $\mathcal R_{\text{min}}=$ 0.2 bit/s.}
\label{fig:Conver_curve}
\end{minipage}
\hspace{.1in} 
\begin{minipage}[t]{0.48\textwidth}
\centering
\includegraphics[width=7.9cm]{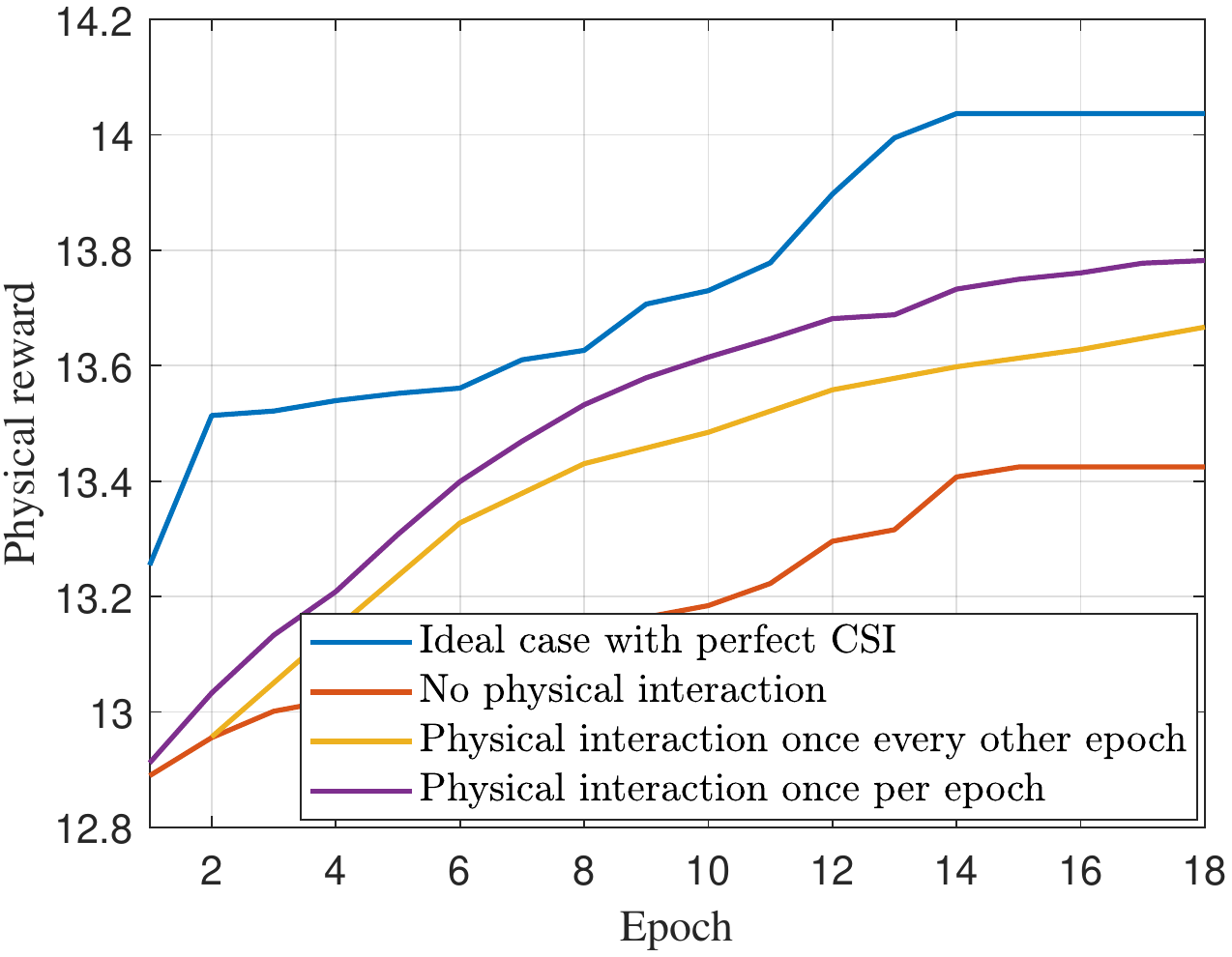}
\caption{The physical rewards of the considered system under different interaction  frequencies between the DT and physical environment. $K=$ 3, $M=$ 4, $B=$ 2, $N=$ 5, $p_{\text{max}}=$ 0.4 W, and $\mathcal R_{\text{min}}=$ 0.2 bit/s.}
\label{fig:phyreward_epoch}
\end{minipage}
\end{figure}

Fig.~\ref{fig:phyreward_epoch} plots the actual rewards obtained by testing the results of our algorithm in the physical environment; c.f. the rewards in the DT in Fig.~\ref{fig:Conver_curve}. As expected, more frequent interactions with the physical environment, e.g., once per epoch, contribute to a higher achievable sum-rate of the proposed algorithm, compared to less frequent interactions of once per two epochs or no interactions with the physical environment at all. 
Nevertheless, the performances of the proposed algorithm exhibit marginal gaps between the different interaction intervals. They are also reasonably close to the ideal case where 
the channel estimations are error-free. 
This validates the effectiveness of the proposed DT-assisted DRL algorithm, especially when the interactions between the server and UE are computationally expensive and incur significant overhead. The use of the DT dramatically reduces the number of direct interactions between the DRL model and the physical environment. In particular, the number of direct interactions is reduced from $Y\times T=40,000$ per epoch in the ideal case (since there are 40,000 steps per epoch) to one per epoch, when the result of the proposed learning model is tested only once in the physical environment at the end of each epoch.

Fig. \ref{fig:learn} shows the impact of different learning rates on the convergence of the proposed algorithm, where the AUA, power control, and RIS beamforming learned by the algorithm with the assistance of the DT are deployed and tested in the physical world once per epoch at the end of every epoch.
By comparing four different learning rates, we observe that the episode reward gradually increases and converges with the increasing number of episodes.
When $r_{\mathrm L}=5\times 10^{-3}$, the episode reward converges fastest to a local optimum. 
An excessively small learning rate accompanied by a small step size can lead the proposed framework to fall into the local optimum, e.g., $r_{\mathrm L}=5\times 10^{-6}$.
The episode reward of the algorithm converges faster under $r_{\mathrm L}=5\times 10^{-4}$ than it does under $r_{\mathrm L}=5\times 10^{-5}$, with comparatively larger spikes.
To this end, it is reasonable to select $r_{\mathrm L}=5\times 10^{-5}$ to balance convergence speed and reward.
}

\begin{figure}[t]
\centering
\begin{minipage}[t]{0.48\textwidth}
\centering
\includegraphics[width=7.9cm]{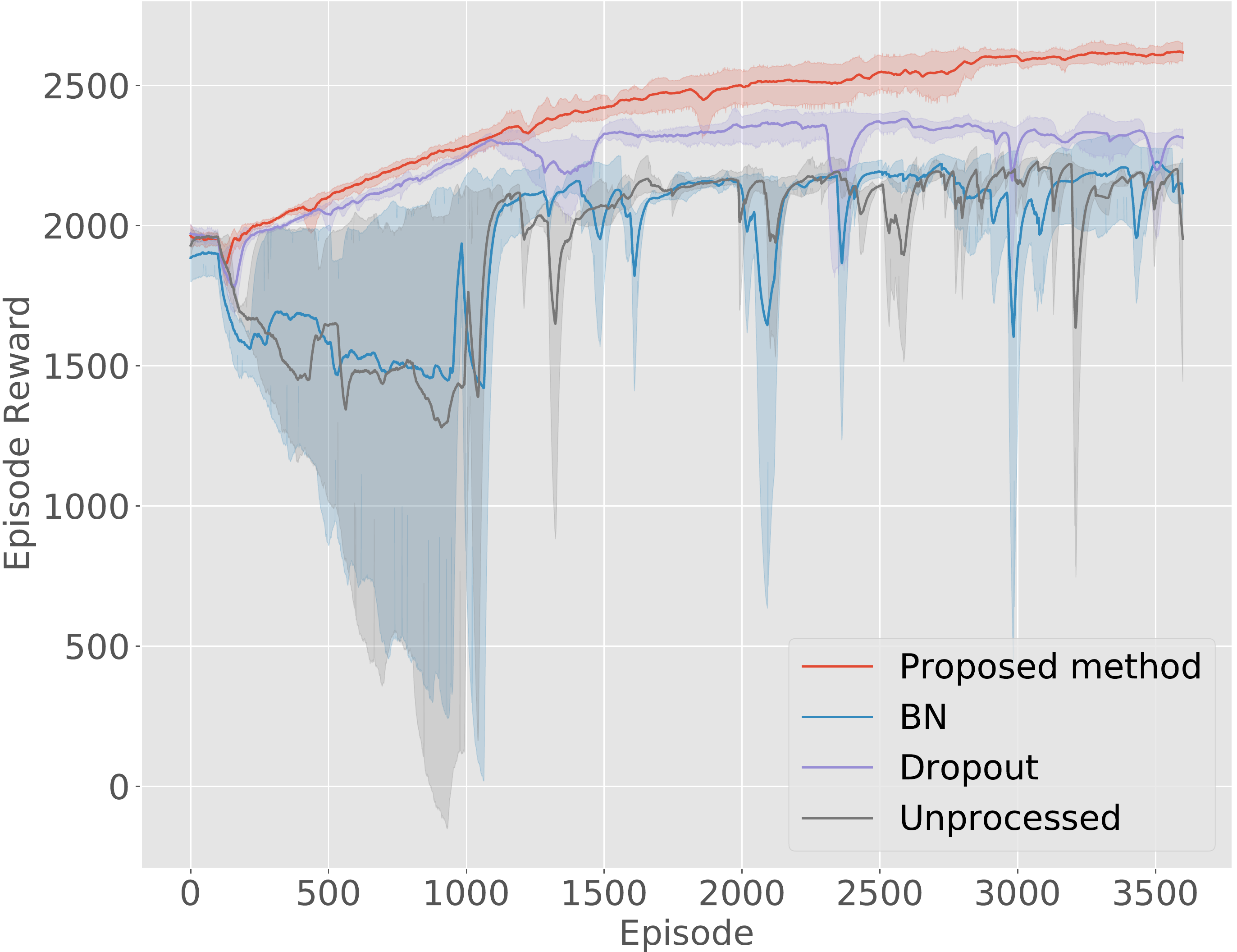}
\caption{{\color{black}The episode reward comparison of different tricks for the neural network. ($K= 3$, $M= 4$, $B= 2$, $N= 5$, $p_{\text{max}}= 0.4$ W and $\mathcal R_{\text{min}}= 0.2$ bit/s.)}}
\label{fig:trick}
\end{minipage}
\hspace{.1in} 
\begin{minipage}[t]{0.48\textwidth}
\centering
\includegraphics[width=7.9cm]{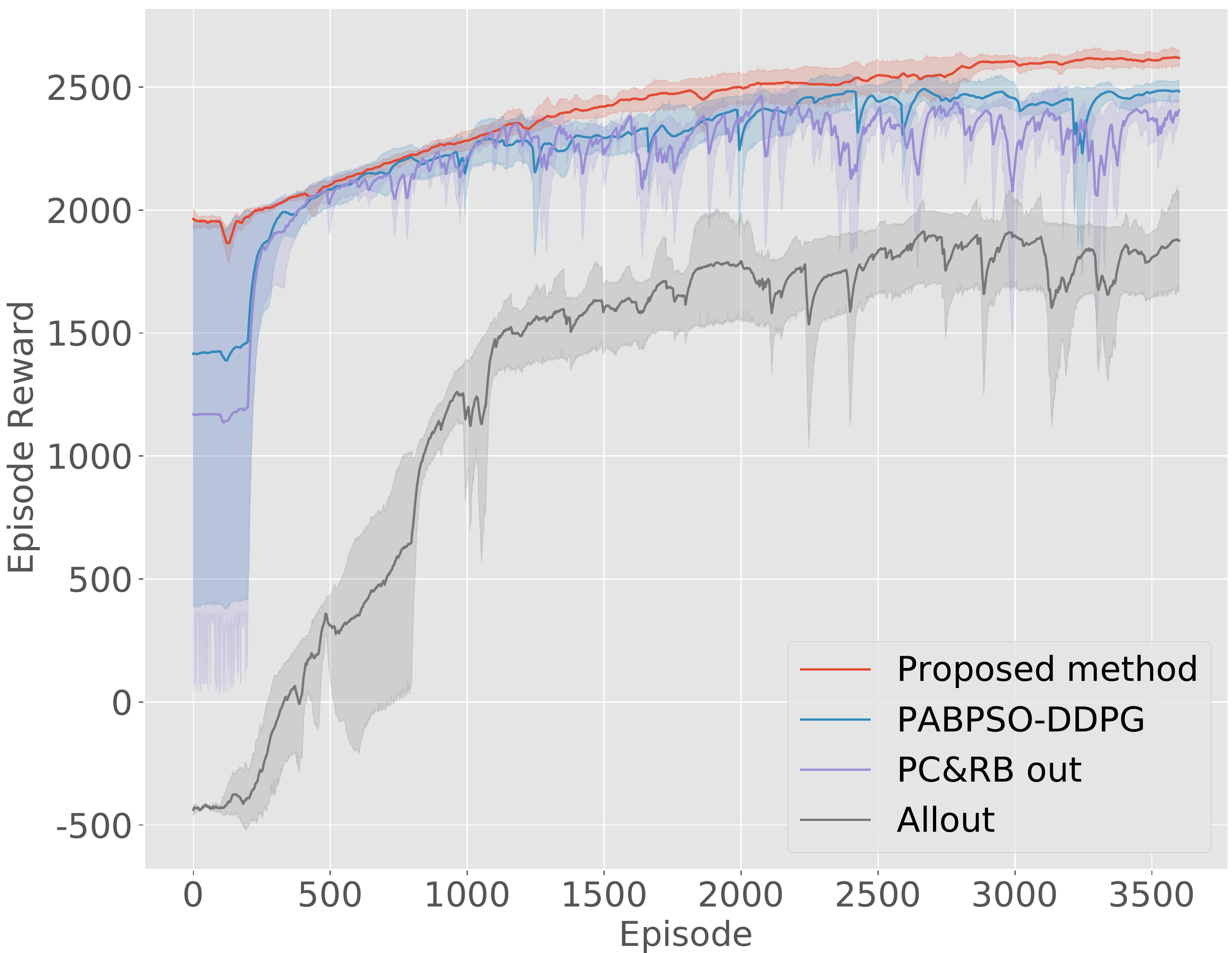}
\caption{{\color{black}The episode reward comparison of different algorithms. ($K=$ 3, $M=$ 4, $B=$ 2, $N=$ 5, $p_{\text{max}}=$ 0.4 W and $\mathcal R_{\text{min}}=$ 0.2 bit/s.)}}
\label{fig:suanfa}
\end{minipage}
\end{figure}

\begin{figure}[t]
\centering
\begin{minipage}[t]{0.47\textwidth}
\centering
\includegraphics[width=7.7cm]{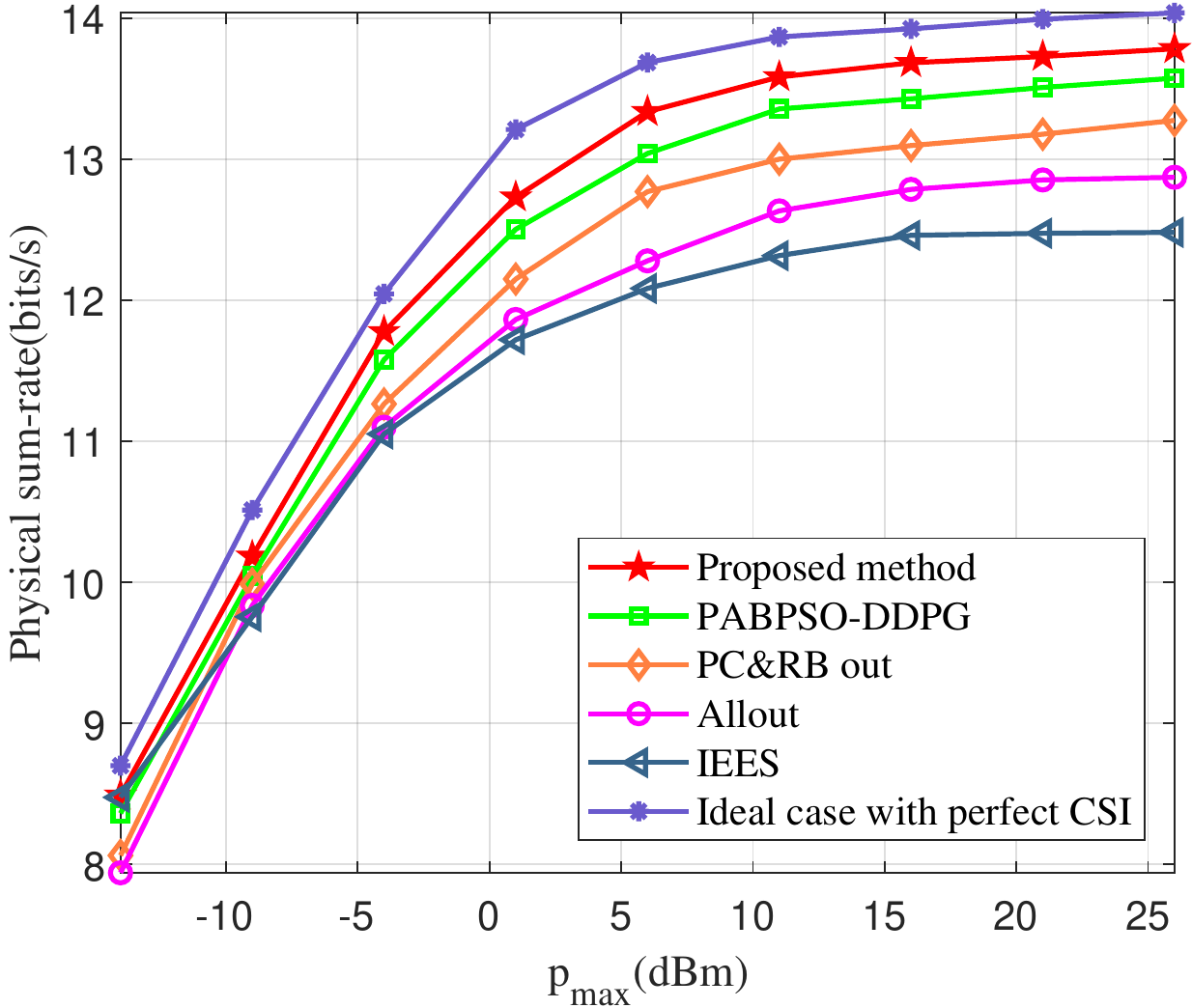}
\caption{{\color{black}The sum-rate comparison of different methods under different maximum transmit power. ($K=$ 3, $M=$ 4, $B=$ 2, $N=$ 5, and $\mathcal R_{\text{min}}=$ 0.2 bit/s.)}}
\label{fig:PMAX}
\end{minipage}
\hspace{.1in} 
\begin{minipage}[t]{0.47\textwidth}
\centering
\includegraphics[width=7.7cm]{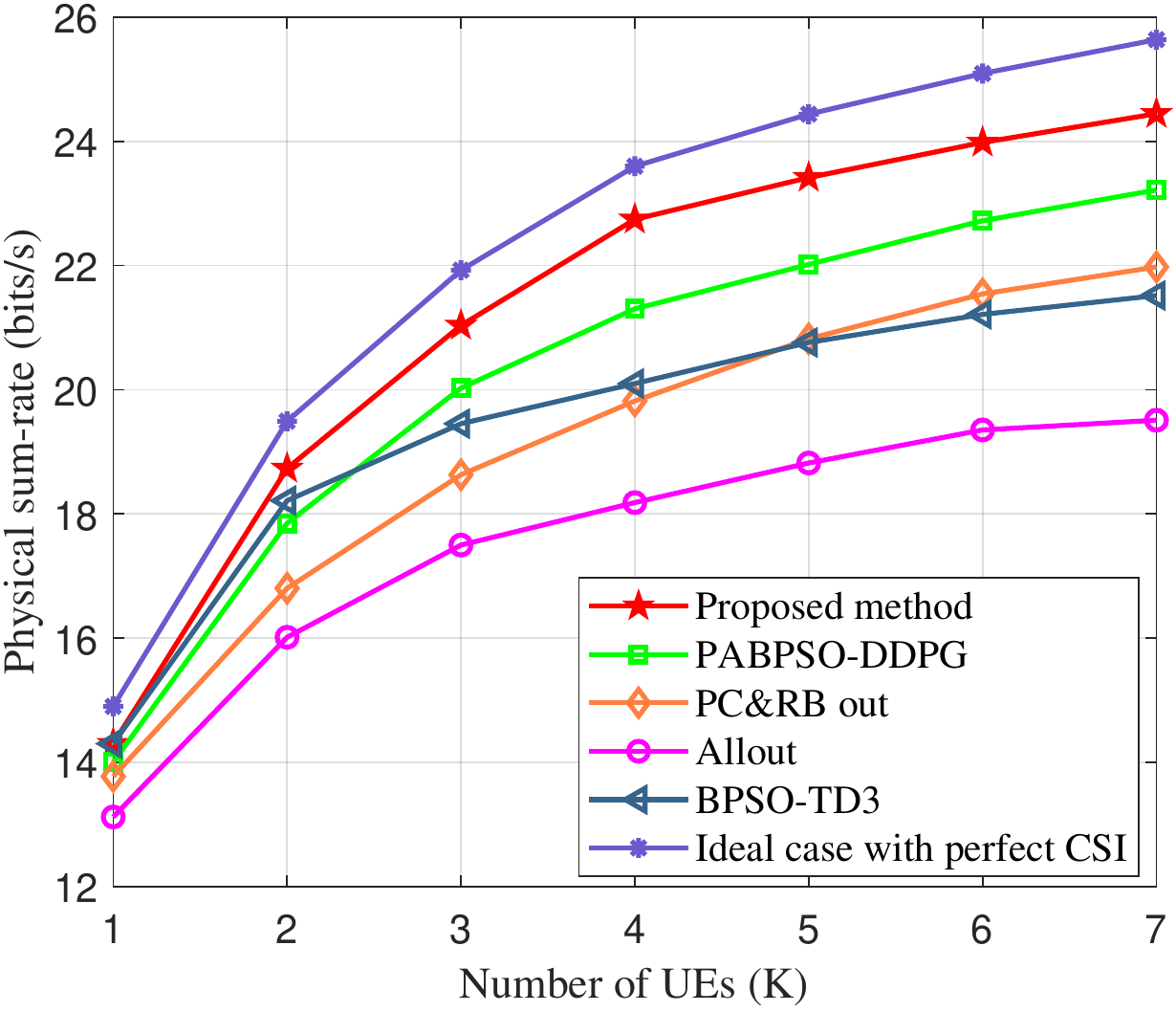}
\caption{{\color{black}The sum-rate comparison of different methods under different $K$. ($M=$ 14, $B=$ 2, $N=$ 5, $p_{\text{max}}=$ 0.04 W and $\mathcal R_{\text{min}}=$ 0.2 bit/s.)}}
\label{fig:UEnum}
\end{minipage}
\end{figure}

\begin{figure}[t]
\centering
\begin{minipage}[t]{0.47\textwidth}
\centering
\includegraphics[width=7.7cm]{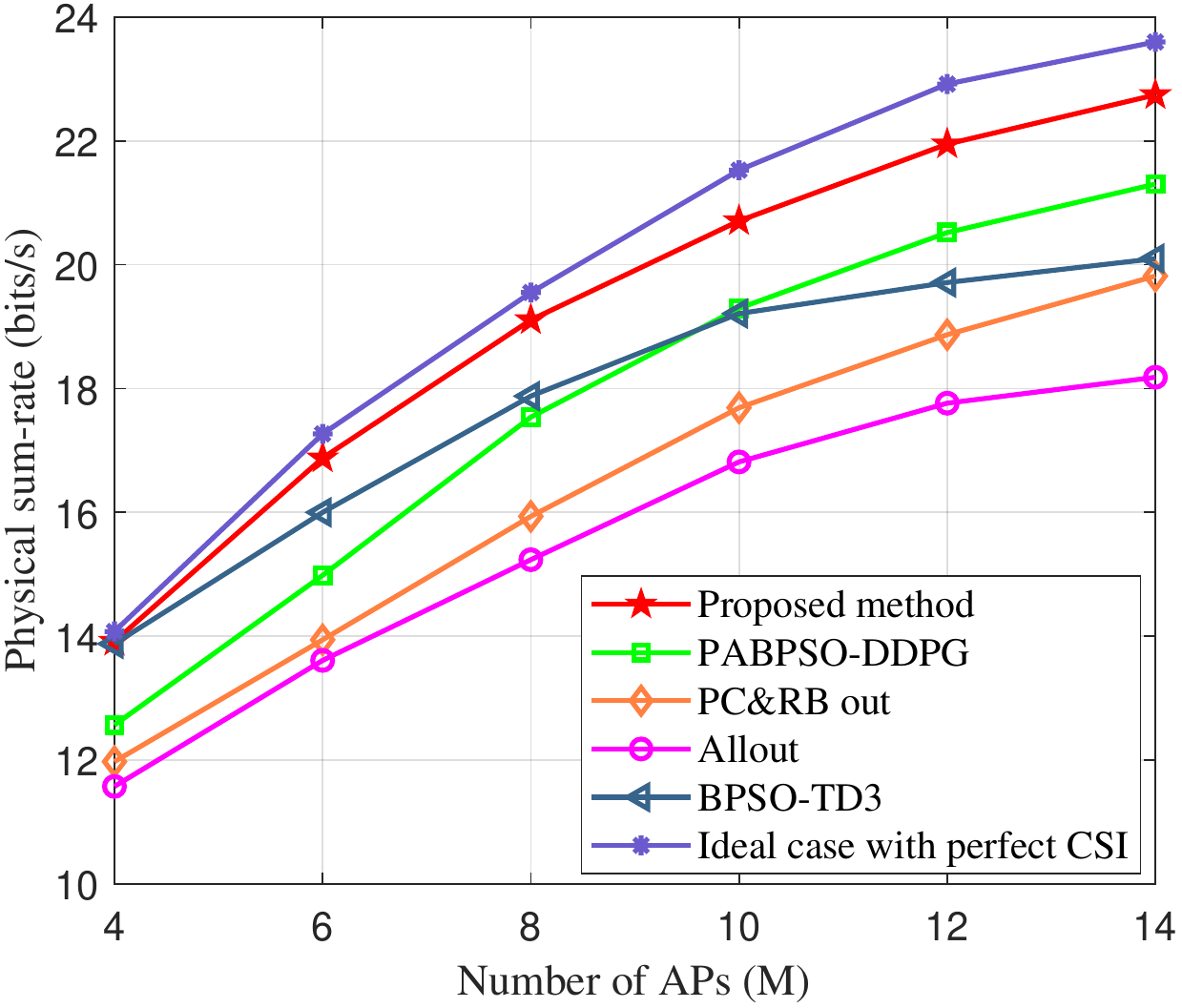}
\caption{{\color{black}The sum-rate comparison of different methods under different $M$. ($K=$ 4, $B=$ 2, $N=$ 5, $p_{\text{max}}=$ 0.04 W and $\mathcal R_{\text{min}}=$ 0.2 bit/s.)}}
\label{fig:APnum}
\end{minipage}
\hspace{.1in} 
\begin{minipage}[t]{0.47\textwidth}
\centering
\includegraphics[width=7.7cm]{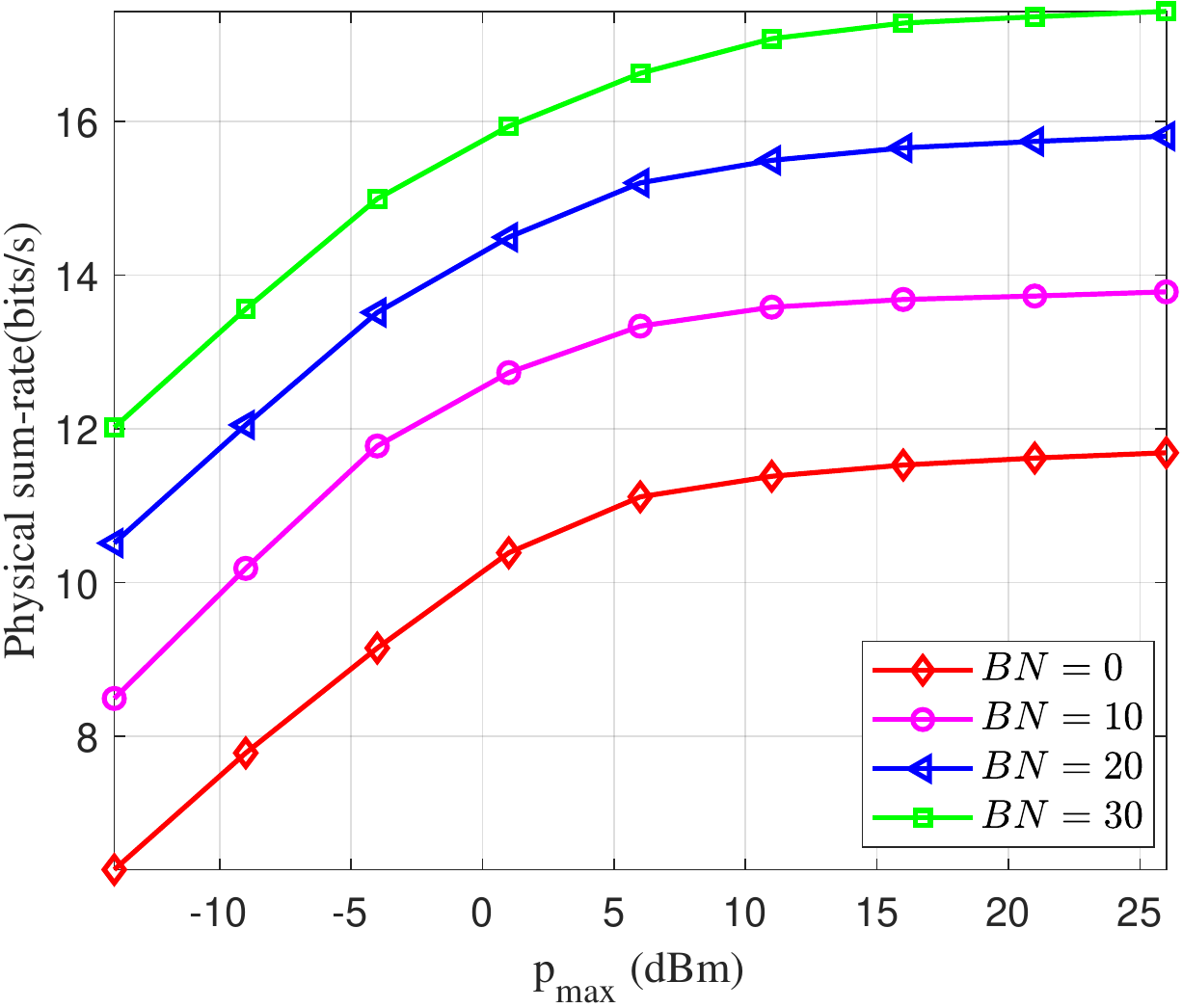}
\caption{The sum-rate comparison of different $BN$ under the different $p_\text{max}$. ($K=$ 3, $M=$ 4, $B=$ 2 and $\mathcal R_{\text{min}}=$ 0.2 bit/s.)}
\label{fig:N_P}
\end{minipage}
\end{figure}

{\color{black} 
Fig.~\ref{fig:trick} evaluates the episode rewards of the PCRB module under different configurations of the TD3 models, including the proposed ``TD3 with max-abs state normalization'', TD3 with batch normalization (BN)~\cite{ioffe2015batch}, TD3 with dropout (with the typical default dropout rate of 0.5)~\cite{hinton2012improving}, and the original TD3 with ``unprocessed'' input state data and no dropout~\cite{fujimoto2018addressing}.
The AUA, power control, and RIS beamforming learned by the algorithms with the assistance of the DT are deployed and tested in the physical world once per epoch at the end of every epoch in this figure and the figures that follow, i.e., Figs.~\ref{fig:suanfa}--\ref{fig:N_P}.
It is observed in Fig.~\ref{fig:trick} that the PCRB module converges the best under state normalization, followed by dropout. The PCRB module performs the worse under BN and ``unprocessed''. The reason is that the value of the unprocessed input data can vary dramatically, some data features may not be effectively learned, and the learning curve concerning the unprocessed data fails to converge.
While BN has the potential to speed up training for stable training data, the varying mean and variance of the training data render BN inapplicable.
Dropout can help the TD3 models quickly converge by randomly cropping some neurons, but the reward can fall into a local optimum  because the dropout of 50\% neurons could be excessive and detrimental to the TD3 models.
}

{\color{black}
Fig.~\ref{fig:suanfa} plots the episode rewards of the PCRB module under different algorithms, including the proposed algorithm, PC\&RB out~\cite{9110869}, Allout~\cite{9345106}, and PABPSO-DDPG~\cite{2015Continuous}. 
It is observed that our algorithm performs the best in terms of convergence, followed by PABPSO-DDPG and PC\&RB out, with the Allout performing the worst.
The proposed algorithm and the PABPSO-DDPG converge more stably than the PC\&RB out and Allout, since the action spaces of the PC\&RB out and Allout are large and unevenly distributed.
It is also observed that the TD3 model exhibits prominent gains over the DDPG model because the DDPG model is prone to overestimate the Q-values in the critic-network and the agents are likely to be trapped at a local optimum due to the accumulated estimation errors.
}

Figs.~\ref{fig:PMAX},~\ref{fig:UEnum}, and~\ref{fig:APnum} study the impacts of the maximum transmit power, $p_{\max}$, and the numbers of UEs and APs, $K$ and $M$, on the physically achievable sum-rate of the proposed algorithm, respectively.
For comparison purposes, we also plot the ideal case with perfect CSI, and the potential alternatives to the proposed algorithm, including PABPSO-DDPG, PC\&RB out, Allout, and IEES~\cite{9359649}.
IEES combines implicit enumeration (IE) and ES, where the transmit power of a UE and the phase shift of an RIS element each are evenly discretized into six levels. It operates based on the environment information recorded at the last interaction of the DT with the physical environment. IE is responsible for enumerating feasible solutions for the AUA. ES searches all possible solutions for power control and RIS beamforming.

It is observed that the proposed algorithm consistently outperforms the rest of the algorithms, and close fairly close to the ideal case.
The IEES performs the worst due to discretization, while all other algorithms support continuous decisions.
Moreover, the growth of the sum-rates slows down as $p_{\text{max}}$ and/or $K$ further increases, resulting from increasingly intensive inter-user interference. 
The figures demonstrate the scalability of the proposed algorithm.

{\color{black}
Fig.~\ref{fig:N_P} assesses the impact of the RISs on the physically achievable sum-rate of the proposed algorithm. It is observed that the sum-rate is significantly higher in the presence of the RISs.
Specifically, the RISs introduce indirect links and provide additional degrees of freedom to enhance useful signal strength and mitigate by adjusting the reflection coefficients of the RISs.
More elements at the RISs contribute to the increase of the sum-rate. 
}

\section{CONCLUSION}\label{section:con}
In this paper, we proposed a new RIS-assisted, uplink UCCF system, maximized its sum-rate, and satisfied its QoS by jointly optimizing the AUA, power control, and RIS beamforming with the aid of a DT.
This was done by transforming this multivariate joint optimization problem into a few interdependent subtasks and proposing a new learning-based framework to separately conduct the subtasks.
By developing a new PABPSO algorithm, the subtask of AUA was efficiently accomplished.
By designing a TD3 model with new and refined state pre-processing layers, the subtasks of power control and RIS beamforming were executed.
The proposed learning framework was trained with the coordination of the DT, thereby dramatically cutting off system overhead.
Simulations corroborated that the RIS-assisted UCCF system can significantly outperform its non-RIS counterparts in sum-rate.
Our DT-aided learning approach provides marked increases in both sum-rate and convergence, compared to its potential alternatives.

\bibliographystyle{IEEEtran}
\bibliography{citations}

\begin{thebibliography}{10}
\providecommand{\url}[1]{#1}
\csname url@samestyle\endcsname
\providecommand{\newblock}{\relax}
\providecommand{\bibinfo}[2]{#2}
\providecommand{\BIBentrySTDinterwordspacing}{\spaceskip=0pt\relax}
\providecommand{\BIBentryALTinterwordstretchfactor}{4}
\providecommand{\BIBentryALTinterwordspacing}{\spaceskip=\fontdimen2\font plus
\BIBentryALTinterwordstretchfactor\fontdimen3\font minus
  \fontdimen4\font\relax}
\providecommand{\BIBforeignlanguage}[2]{{%
\expandafter\ifx\csname l@#1\endcsname\relax
\typeout{** WARNING: IEEEtran.bst: No hyphenation pattern has been}%
\typeout{** loaded for the language `#1'. Using the pattern for}%
\typeout{** the default language instead.}%
\else
\language=\csname l@#1\endcsname
\fi
#2}}
\providecommand{\BIBdecl}{\relax}
\BIBdecl

\bibitem{8097026}
H.~Q. Ngo, L.-N. Tran, T.~Q. Duong, M.~Matthaiou, and E.~G. Larsson, ``On the
  total energy efficiency of cell-free massive {MIMO},'' \emph{IEEE Trans.
  Green Commun. Netw.}, vol.~2, no.~1, pp. 25--39, Jan. 2018.

\bibitem{7227028}
H.~Q. Ngo \emph{et~al.}, ``Cell-free massive {MIMO}: Uniformly great service
  for everyone,'' in \emph{Proc. IEEE Int. Workshop Signal Proc. Adv. Wirel.
  Commun. (SPAWC)}, Stockholm, Sweden, Jun. 2015, pp. 201--205.

\bibitem{7827017}
H.~Q. Ngo, A.~Ashikhmin, H.~Yang, E.~G. Larsson, and T.~L. Marzetta,
  ``Cell-free massive {MIMO} versus small cells,'' \emph{IEEE Trans. Wirel.
  Commun.}, vol.~16, no.~3, pp. 1834--1850, Mar. 2017.

\bibitem{8886730}
P.~Liu, K.~Luo, D.~Chen, and T.~Jiang, ``Spectral efficiency analysis of
  cell-free massive {MIMO} systems with zero-forcing detector,'' \emph{IEEE
  Trans. Wirel. Commun.}, vol.~19, no.~2, pp. 795--807, Feb. 2020.

\bibitem{8000355}
S.~Buzzi and C.~D'Andrea, ``Cell-free massive {MIMO}: User-centric approach,''
  \emph{IEEE Wirel. Commun. Lett.}, vol.~6, no.~6, pp. 706--709, 2017.

\bibitem{8901451}
S.~Buzzi, C.~D'Andrea, A.~Zappone, and C.~D’Elia, ``User-centric 5{G}
  cellular networks: Resource allocation and comparison with the cell-free
  massive {MIMO} approach,'' \emph{IEEE Trans. Wirel. Commun.}, vol.~19, no.~2,
  pp. 1250--1264, Feb. 2020.

\bibitem{8910627}
Q.~Wu and R.~Zhang, ``Towards smart and reconfigurable environment: Intelligent
  reflecting surface aided wirel. network,'' \emph{IEEE Commun. Mag.}, vol.~58,
  no.~1, pp. 106--112, Jan. 2020.

\bibitem{9459505}
Z.~Zhang and L.~Dai, ``A joint precoding framework for wideband reconfigurable
  intelligent surface-aided cell-free network,'' \emph{IEEE Trans. Signal
  Process.}, vol.~69, pp. 4085--4101, Jun. 2021.

\bibitem{9115725}
S.~Abeywickrama, R.~Zhang, Q.~Wu, and C.~Yuen, ``Intelligent reflecting
  surface: Practical phase shift model and beamforming optimization,''
  \emph{IEEE Trans. Commun.}, vol.~68, no.~9, pp. 5849--5863, Sept. 2020.

\bibitem{9363171}
Q.~N. Le, V.-D. Nguyen, O.~A. Dobre, and R.~Zhao, ``Energy efficiency
  maximization in {RIS}-aided cell-free network with limited backhaul,''
  \emph{IEEE Commun. Lett.}, vol.~25, no.~6, pp. 1974--1978, Jun. 2021.

\bibitem{8790780}
C.~He, Y.~Hu, Y.~Chen, and B.~Zeng, ``Joint power allocation and channel
  assignment for {NOMA} with deep reinforcement learning,'' \emph{IEEE J. Sel.
  Areas Commun.}, vol.~37, no.~10, pp. 2200--2210, Oct. 2019.

\bibitem{8714026}
N.~C. Luong, D.~T. Hoang, S.~Gong, D.~Niyato, P.~Wang, Y.-C. Liang, and D.~I.
  Kim, ``Applications of deep reinforcement learning in communications and
  networking: A survey,'' \emph{IEEE Commun. Surv. Tut.}, vol.~21, no.~4, pp.
  3133--3174, Fourthquarter 2019.

\bibitem{8822494}
C.~Gehrmann and M.~Gunnarsson, ``A digital twin based industrial automation and
  control system security architecture,'' \emph{IEEE Trans. Ind. Informat.},
  vol.~16, no.~1, pp. 669--680, Jan. 2020.

\bibitem{9447819}
T.~Liu, L.~Tang, W.~Wang, Q.~Chen, and X.~Zeng, ``Digital-twin-assisted task
  offloading based on edge collaboration in the digital twin edge network,''
  \emph{IEEE Internet Things J.}, vol.~9, no.~2, pp. 1427--1444, Jan. 2022.

\bibitem{8972134}
F.~Pires, A.~Cachada, J.~Barbosa, A.~P. Moreira, and P.~Leitão, ``Digital twin
  in industry 4.0: Technologies, applications and challenges,'' in \emph{2019
  IEEE 17th Int. Conf. Ind. Informat. (INDIN)}, vol.~1, Helsinki, Finland, Jul.
  2019, pp. 721--726.

\bibitem{luan2021paradigm}
T.~H. Luan, R.~Liu, L.~Gao, R.~Li, and H.~Zhou, ``The paradigm of digital twin
  communications,'' \emph{arXiv preprint arXiv:2105.07182}, 2021.

\bibitem{9351542}
W.~Sun, P.~Wang, N.~Xu, G.~Wang, and Y.~Zhang, ``Dynamic digital twin and
  distributed incentives for resource allocation in aerial-assisted internet of
  vehicles,'' \emph{IEEE Internet Things J.}, vol.~9, no.~8, pp. 5839--5852,
  Apr. 2022.

\bibitem{sheen2020digital}
B.~Sheen, J.~Yang, X.~Feng, and M.~M.~U. Chowdhury, ``A digital twin for
  reconfigurable intelligent surface assisted wireless communication,''
  \emph{arXiv preprint arXiv:2009.00454}, 2020.

\bibitem{9420037}
J.~Deng, Q.~Zheng, G.~Liu, J.~Bai, K.~Tian, C.~Sun, Y.~Yan, and Y.~Liu, ``A
  digital twin approach for self-optimization of mobile networks,'' in
  \emph{2021 IEEE Wirel. Commun. Netw. Conf. Workshops (WCNCW)}, Nanjing,
  China, Mar. 2021, pp. 1--6.

\bibitem{lv2021beyond}
Z.~Lv, D.~Chen, H.~Feng, R.~Lou, and H.~Wang, ``Beyond {5G} for digital twins
  of {UAV}s,'' \emph{Comput. Netw.}, vol. 197, p. 108366, 2021.

\bibitem{8676377}
M.~Alonzo, S.~Buzzi, A.~Zappone, and C.~D’Elia, ``Energy-efficient power
  control in cell-free and user-centric massive {MIMO} at millimeter wave,''
  \emph{IEEE Trans. Green Commun. Netw.}, vol.~3, no.~3, pp. 651--663, Sept.
  2019.

\bibitem{7676391}
A.~Liu and V.~K.~N. Lau, ``Joint {BS}-user association, power allocation, and
  user-side interference cancellation in cell-free heterogeneous networks,''
  \emph{IEEE Trans. Signal Process.}, vol.~65, no.~2, pp. 335--345, Jan. 2017.

\bibitem{9443536}
C.~D'Andrea and E.~G. Larsson, ``User association in scalable cell-free massive
  {MIMO} systems,'' in \emph{Proc. Asilomar Conf. Signals, Sys. Comp. (ACSSC)},
  Nov. 2020, pp. 826--830.

\bibitem{9448858}
K.~Liu and Z.~Zhang, ``On the energy-efficiency fairness of reconfigurable
  intelligent surface-aided cell-free network,'' in \emph{Proc. IEEE 93rd Veh.
  Technol. Conf. (VTC-Spring)}, Apr. 2021, pp. 1--6.

\bibitem{9322151}
M.~Bashar, K.~Cumanan, A.~G. Burr, P.~Xiao, and M.~Di~Renzo, ``On the
  performance of reconfigurable intelligent surface-aided cell-free massive
  {MIMO} uplink,'' in \emph{Proc. IEEE Global Commun. Conf. (GLOBECOM)}, Dec.
  2020, pp. 1--6.

\bibitem{9298843}
S.~Huang, Y.~Ye, M.~Xiao, H.~V. Poor, and M.~Skoglund, ``Decentralized
  beamforming design for intelligent reflecting surface-enhanced cell-free
  networks,'' \emph{IEEE Wirel. Commun. Lett.}, vol.~10, no.~3, pp. 673--677,
  Mar. 2021.

\bibitem{9352948}
Y.~Zhang, B.~Di, H.~Zhang, J.~Lin, C.~Xu, D.~Zhang, Y.~Li, and L.~Song,
  ``Beyond cell-free {MIMO}: Energy efficient reconfigurable intelligent
  surface aided cell-free {MIMO} communications,'' \emph{IEEE Trans. Cogn.
  Commun. Netw.}, vol.~7, no.~2, pp. 412--426, Jun. 2021.

\bibitem{9483914}
Y.~Zhao, I.~G. Niemegeers, and S.~M.~H. De~Groot, ``Dynamic power allocation
  for cell-free massive {MIMO}: Deep reinforcement learning methods,''
  \emph{IEEE Access}, vol.~9, pp. 102\,953--102\,965, 2021.

\bibitem{2020distributedDDPG}
F.~Fredj, Y.~Al-Eryani, S.~Maghsudi, M.~Akrout, and E.~Hossain, ``Distributed
  uplink beamforming in cell-free networks using deep reinforcement learning,''
  \emph{arXiv preprint arXiv:2006.15138}, 2020.

\bibitem{9174775}
Y.~Al-Eryani, M.~Akrout, and E.~Hossain, ``Multiple access in cell-free
  networks: Outage performance, dynamic clustering, and deep reinforcement
  learning-based design,'' \emph{IEEE J. Sel. Areas Commun.}, vol.~39, no.~4,
  pp. 1028--1042, Apr. 2021.

\bibitem{8968350}
K.~Feng, Q.~Wang, X.~Li, and C.-K. Wen, ``Deep reinforcement learning based
  intelligent reflecting surface optimization for {MISO} communication
  systems,'' \emph{IEEE Wirel. Commun. Lett.}, vol.~9, no.~5, pp. 745--749, May
  2020.

\bibitem{9110869}
C.~Huang, R.~Mo, and C.~Yuen, ``Reconfigurable intelligent surface assisted
  multiuser {MISO} systems exploiting deep reinforcement learning,'' \emph{IEEE
  J. Sel. Areas Commun.}, vol.~38, no.~8, pp. 1839--1850, Aug. 2020.

\bibitem{9473585}
J.~Kim, S.~Hosseinalipour, T.~Kim, D.~J. Love, and C.~G. Brinton,
  ``Multi-{IRS}-assisted multi-cell uplink {MIMO} communications under
  imperfect {CSI}: A deep reinforcement learning approach,'' in \emph{Proc.
  IEEE Int. Conf. Commun. Workshops (ICC Workshops)}, Jul. 2021, pp. 1--7.

\bibitem{10.1117/12.2618612}
\BIBentryALTinterwordspacing
M.~McManus, Z.~Guan, N.~Mastronarde, and S.~Zou, ``{On the source-to-target gap
  of robust double deep Q-learning in digital twin-enabled wireless
  networks},'' in \emph{{Big Data IV}: Learning, Analytics, and Applications},
  F.~Ahmad, P.~P. Markopoulos, and B.~Ouyang, Eds., vol. 12097, International
  Society for Optics and Photonics.\hskip 1em plus 0.5em minus 0.4em\relax
  SPIE, May 2022, p. 1209706. [Online]. Available:
  \url{https://doi.org/10.1117/12.2618612}
\BIBentrySTDinterwordspacing

\bibitem{9732214}
G.~Zhou, C.~Pan, H.~Ren, P.~Popovski, and A.~L. Swindlehurst, ``Channel
  estimation for {RIS}-aided multiuser millimeter-wave systems,'' \emph{IEEE
  Trans. Signal Process.}, vol.~70, pp. 1478--1492, Mar. 2022.

\bibitem{8645336}
H.~Q. Ngo, H.~Tataria, M.~Matthaiou, S.~Jin, and E.~G. Larsson, ``On the
  performance of cell-free massive {MIMO} in {R}icean fading,'' in \emph{Proc.
  IEEE Conf. Rec. Asilomar Conf. Signals, Sys. Comp. (ACSSC)}, Nov. 2018, pp.
  980--984.

\bibitem{dai2022two}
J.~Dai, J.~Ge, K.~Zhi, C.~Pan, Z.~Zhang, J.~Wang, and X.~You, ``Two-timescale
  transmission design for {RIS}-aided cell-free massive {MIMO} systems,''
  \emph{arXiv preprint arXiv:2210.08514}, 2022.

\bibitem{Li2019Energy}
K.~Li, R.~C. Voicu, S.~S. Kanhere, W.~Ni, and E.~Tovar, ``Energy efficient
  legitimate wireless surveillance of uav communications,'' \emph{IEEE Trans.
  Veh. Tech.}, vol.~68, no.~3, pp. 2283--2293, 2019.

\bibitem{ding2020deep}
H.~Ding, F.~Zhao, J.~Tian, D.~Li, and H.~Zhang, ``A deep reinforcement learning
  for user association and power control in heterogeneous networks,'' \emph{Ad
  Hoc Netw.}, vol. 102, p. 102069, May 2020.

\bibitem{Wang2015VANET}
H.~Wang, R.~P. Liu, W.~Ni, W.~Chen, and I.~B. Collings, ``Vanet modeling and
  clustering design under practical traffic, channel and mobility conditions,''
  \emph{IEEE Trans. Commun.}, vol.~63, no.~3, pp. 870--881, 2015.

\bibitem{Li2019On}
K.~Li, W.~Ni, E.~Tovar, and A.~Jamalipour, ``On-board deep q-network for
  uav-assisted online power transfer and data collection,'' \emph{IEEE Trans.
  Veh. Tech.}, vol.~68, no.~12, pp. 12\,215--12\,226, 2019.

\bibitem{637339}
J.~Kennedy and R.~Eberhart, ``A discrete binary version of the particle swarm
  algorithm,'' in \emph{Proc. IEEE Int. Conf. Syst., Man, Cybern.: Comput.
  Cybern. Simulation (ICSMC)}, vol.~5, Oct. 1997, pp. 4104--4108 vol. 5.

\bibitem{Li2016Energy}
K.~Li \emph{et~al.}, ``Energy-efficient cooperative relaying for unmanned
  aerial vehicles,'' \emph{IEEE Trans. Mobile Comput.}, vol.~15, no.~6, pp.
  1377--1386, 2016.

\bibitem{9145209}
Z.~Li, M.~Chen, K.~Wang, C.~Pan, N.~Huang, and Y.~Hu, ``Parallel deep
  reinforcement learning based online user association optimization in
  heterogeneous networks,'' in \emph{Proc. IEEE Int. Conf. Commun. Workshops
  (ICC Workshops)}, Jun. 2020, pp. 1--6.

\bibitem{9111671}
O.~Llerena-Pizarro, N.~Proenza-Perez, C.~E. Tuna, and J.~L. Silveira, ``A
  {PSO}-{BPSO} technique for hybrid power generation system sizing,''
  \emph{IEEE Lat. Am. Trans.}, vol.~18, no.~08, pp. 1362--1370, Aug. 2020.

\bibitem{xu2019targeted}
N.~Xu, G.~Zheng, K.~Xu, Y.~Zhu, and Z.~Li, ``Targeted knowledge transfer for
  learning traffic signal plans,'' in \emph{Pacific-Asia Conf. Knowl. Discov.
  Data Min.}\hskip 1em plus 0.5em minus 0.4em\relax Springer, Mar. 2019, pp.
  175--187.

\bibitem{sutton2018reinforcement}
R.~S. Sutton and A.~G. Barto, \emph{Reinforcement learning: An
  introduction}.\hskip 1em plus 0.5em minus 0.4em\relax MIT press, 2018.

\bibitem{2015Continuous}
T.~P. Lillicrap, J.~J. Hunt, A.~Pritzel, N.~Heess, T.~Erez, Y.~Tassa,
  D.~Silver, and D.~Wierstra, ``Continuous control with deep reinforcement
  learning,'' \emph{arXiv preprint arXiv:1509.02971}, Sept. 2015.

\bibitem{fujimoto2018addressing}
S.~Fujimoto, H.~Hoof, and D.~Meger, ``Addressing function approximation error
  in actor-critic methods,'' in \emph{Proc. 35th Int. Conf. Mach. Learn.
  (ICML)}.\hskip 1em plus 0.5em minus 0.4em\relax PMLR, Jul. 2018, pp.
  1587--1596.

\bibitem{articlePrognosisModel}
S.~Amarnath, M.~Selvamani, and V.~Varadarajan, ``Prognosis model for
  gestational diabetes using machine learning techniques,'' \emph{Sens.
  Mater.}, vol.~33, p. 3011, May. 2021.

\bibitem{9575181}
R.~Zhang, K.~Xiong, Y.~Lu, B.~Gao, P.~Fan, and K.~B. Letaief, ``Joint
  coordinated beamforming and power splitting ratio optimization in {MU-MISO
  SWIPT}-enabled {H}et{N}ets: A multi-agent {DDQN}-based approach,'' \emph{IEEE
  J. Sel. Areas Commun.}, vol.~40, no.~2, pp. 677--693, Feb. 2022.

\bibitem{9345106}
Z.~Lyu, C.~Ren, and L.~Qiu, ``Movement and communication co-design in
  multi-{UAV} enabled wireless systems via {DRL},'' in \emph{2020 IEEE 6th
  International Conference on Computer and Communications (ICCC)}, 2020, pp.
  220--226.

\bibitem{ioffe2015batch}
S.~Ioffe and C.~Szegedy, ``Batch normalization: Accelerating deep network
  training by reducing internal covariate shift,'' in \emph{Proc. 32nd Int.
  Conf. Mach. Learn. (ICML)}.\hskip 1em plus 0.5em minus 0.4em\relax PMLR,
  2015, pp. 448--456.

\bibitem{hinton2012improving}
G.~E. Hinton, N.~Srivastava, A.~Krizhevsky, I.~Sutskever, and R.~R.
  Salakhutdinov, ``Improving neural networks by preventing co-adaptation of
  feature detectors,'' \emph{arXiv preprint arXiv:1207.0580}, 2012.

\bibitem{9359649}
D.~Xu, ``Proactive eavesdropping over {OFDM}-based bidirectional suspicious
  communication channels,'' \emph{IEEE Wirel. Commun. Lett.}, vol.~10, no.~6,
  pp. 1178--1182, Jun. 2021.

\end{thebibliography}
\end{document}